\documentclass[useAMS,usenatbib]{mn2e}
\pdfoutput=1
\usepackage{amssymb}
\usepackage{graphicx}
\usepackage[usenames,dvipsnames]{color}
\usepackage{verbatim}

\usepackage{times}
\usepackage{mathptmx}

\newcommand{\um}[1]{\,\mathrm{#1}}
\newcommand{\mic}[1]{\,\umu\mathrm{#1}}
\newcommand{\HAng}[3]{#1\!:\!#2\!:\!#3}
\newcommand{\DAng}[3]{#1^\circ\,#2'\,#3''}

\title[Radio emission of two SNRs and four PNe]{Study of the extended radio emission of two supernova remnants and four planetary nebulae associated to MIPSGAL bubbles}
\author[A. Ingallinera et al.]{{\large A. Ingallinera$^{1,2}$\thanks{E-mail:ingallinera@oact.inaf.it}, C. Trigilio$^2$, G. Umana$^2$, P. Leto$^2$, C. Agliozzo$^{3,4}$, C. Buemi$^2$}\\
$^1$Universit\`a di Catania, Dipartimento di Fisica e Astronomia, Via Santa Sofia, 64, 95123 Catania, Italy\\
$^2$INAF-Osservatorio Astrofisico di Catania, Via Santa Sofia 78, 95123 Catania, Italy\\
$^3$Millennium Institute of Astrophysics, Santiago, Chile\\
$^4$Universidad Andr\'es Bello, Avda. Republica 252, Santiago, Chile
}

\begin{document}

\date{Accepted 2014 October 2. Received 2014 September 19; in original form 2014 August 4}

\pagerange{\pageref{firstpage}--\pageref{lastpage}} \pubyear{2014}

\maketitle

\label{firstpage}

\begin{abstract}
We present radio observations of two supernova remnants and four planetary nebulae with the Very Large Array and the Green Bank Telescope. These objects are part of a larger sample of radio sources, discussed in a previous paper, counterpart of the MIPSGAL 24-$\umu$m compact bubbles. For the two supernova remnants we combined the interferometric observations with single-dish data to obtain both a high resolution and a good sensitivity to extended structures. We discuss in detail the entire combination procedure adopted and the reliability of the resulting maps. For one supernova remnant we pose a more stringent upper limit for the flux density of its undetected pulsar, and we also show prominent spectral index spatial variations, probably due to inhomogeneities in the magnetic field and in its ejecta or to an interaction between the supernova shock and molecular clouds. We eventually use the 5-GHz maps of the four planetary nebulae to estimate their distance and their ionized mass.

\end{abstract}


\section{Introduction}
From the visual inspection of the MIPSGAL Legacy Survey \citep{Carey2009} mosaic images, obtained with MIPS\footnote{The Multiband Imaging Photometer for Spitzer.} \citep{Rieke2004} on board of the \textit{Spitzer Space Telescope}, \citet{Mizuno2010} identified at $24\mic{m}$ 428 compact roundish objects presenting diffuse emission. These small ($\lesssim1'$) rings, disks or shells (hereafter denoted as `bubbles') are pervasive throughout the entire Galactic plane in the mid-infrared (IR). The main hypothesis about the nature of the bubbles is that they are different type of evolved stars (planetary nebulae, supernova remnants, Wolf--Rayet stars, asymptotic giant branch stars, etc.). However, currently, only about 30 per cent of the bubbles are classified.

In order to shed light on the nature of the bubbles, we had carried out radio continuum observations with the Karl G.~Jansky Very Large Array (VLA) in 2010 ($C$ band, configuration D) and 2012 ($L$ band, configuration C and CnB), on a subset of 55 bubbles (\citealt{Ingallinera2014}; `Paper I' hereafter). We were able to calculate the radio spectral index for 31 of them, finding that at least the 70 per cent are radio thermal emitters. Among the 55 bubbles observed at $5\um{GHz}$, ten were classified and, mostly, well known objects. We had excluded these objects from the remainder of that work, mainly aimed at characterising unclassified objects.

In this paper we present an analysis of the VLA radio continuum observations described in Paper I for six of these ten classified bubbles, namely two supernova remnants (SNRs) and four planetary nebulae (PNe), addressed to improve the knowledge of their physical properties. For the two SNRs, whose extension is $\sim\!4\um{arcmin}$, we also conducted single-dish observations with the Green Bank Telescope (GBT). The purpose of these observations was to complement VLA maps with the low spatial frequency information provided by the GBT, to achieve both a high resolution and a high reliability in flux density determination. Also two PNe ($\sim\!1\um{arcmin}$) were observed with the GBT to check if undetected extended emission existed and what kind of improvement, if any, the single-dish data would have brought to VLA maps in this case. The GBT data reduction process and the combination method will be discussed thoroughly, since, by now, no unanimously-accepted standard procedures exist.

\section{Observations and data reduction}

\subsection{Interferometric observations and data reduction}
\label{sec:vla_obs}
The interferometric observations and the following data reduction were described in detail in Paper I. Therefore here we will limit to a brief summary.

From the original list of the 428 MIPSGAL bubbles we selected only sources visible at the VLA latitude and with a possible detection in NVSS\footnote{http://www.cv.nrao.edu/nvss/NVSSlist.shtml \citep{Condon1998}.} or in MAGPIS\footnote{http://third.ucllnl.org/gps/catalogs.html \citep{Helfand2006}.}. The final subset consisted of 55 bubbles, that were observed at $5\um{GHz}$ ($C$ band) in March 2010 with the VLA in configuration D. Among these 55 bubbles, 40 were subsequently observed at $1.4\um{GHz}$ ($L$ band) in March and May 2012 in configuration C and CnB. For $C$-band observations, each bubble was observed for slightly less than 10 minutes with a total bandwidth of $256\um{MHz}$. The average synthetic beam size was around $25\times15\um{arcsec}^2$ and the typical map rms $\sim\!100\mic{Jy/beam}$. For $L$-band observations, each bubble was observed for 10 to 20 minutes with a total bandwidth of $1\um{GHz}$. The average beam size was $18\times12\um{arcsec}^2$. The presence of conspicuous radio-frequency interferences limited the sensitivity reached in these observations, and the typical map rms was about $0.5-1\um{mJy/beam}$.

The entire data reduction process was performed using the package \textsc{casa}\footnote{http://casa.nrao.edu \citep{McMullin2007}.}. For all the observations, the bandpass and flux calibrator was 3C286, while for gain calibration, we used a variety of standard calibrators, depending on their distance from the sources (typically within $10^\circ$).

Differently from Paper I, the imaging procedure involved the use of the `multi-scale algorithm' \citep{Cornwell2008}. While the standard Clark implementation \citep{Clark1980} of the CLEAN algorithm \citep{Hogbom1974} assumes that the sky is a collection of point sources on an empty background, the multi-scale CLEAN permits the presence  of sources of many different sizes and scales. This algorithm is therefore extremely useful when an extended source is to be imaged. In particular we defined three different scales for source extensions: 0 pixel to search for point sources (equivalent to a normal CLEAN and necessary to properly consider nearby point sources), 5 pixels (about the beam dimension) and 15 pixels.

In Table \ref{tab:GBTsample} we list all the sources studied in this paper, indicating their main designation, the \citet{Mizuno2010} name ([MGE]) and whether we observed them with the VLA and with the GBT.

\begin{table*}
\begin{center}
\begin{tabular}{lcccccc}\hline\hline
Designation & [MGE] & Right ascension & Declination & \multicolumn{3}{c}{Observations}\\
&& (J2000) & (J2000) & VLA $5\um{GHz}$ & VLA $1.4\um{GHz}$ & GBT\\\hline
SNR G011.2-00.3 & 011.1805-00.3471 & $\HAng{18}{11}{28.9}$ & $\DAng{-19}{25}{29}$ & yes & no & yes\\ 
SNR G027.4+00.0 & 027.3891-00.0079 & $\HAng{18}{41}{19.9}$ & $\DAng{-04}{56}{06}$ & yes & yes & yes\\ 
PN G029.0+00.4 & 029.0784+00.4545	 & $\HAng{18}{42}{46.8}$ & $\DAng{-03}{13}{17}$ & yes & yes & yes\\ 
NGC 6842 & 065.9141+00.5966 & $\HAng{19}{55}{02.4}$ & $\DAng{+29}{17}{20}$ & yes & no & yes\\ 
PN G040.3-00.4 & 040.3704-00.4750 & $\HAng{19}{06}{45.8}$ & $\DAng{+06}{23}{53}$ & yes & yes & no\\ 
PN G031.9-00.3 & 031.9075-00.3087 & $\HAng{18}{50}{40.1}$ & $\DAng{-01}{03}{09}$ & yes & yes & no\\ 
\hline

\end{tabular}
\label{tab:GBTsample}
\end{center}
\caption{List of the sources studied in this work.}
\end{table*}

\subsection{Missing low spatial frequencies}
An interferometer is an instrument capable to measure the visibility function $V(u,v)$ that, under certain circumstances, is the Fourier transform of the true sky brightness (e.g. \citealt{Clark1999}). However, the instrument is able to measure $V(u,v)$ only on a discrete set of $(u,v)$ pairs, given by the interferometer baseline projections on the $uv$ plane. Even if the Earth rotation is exploited, the $uv$ plane coverage will be always incomplete. In particular there will be a minimum value for $\sqrt{u^2+v^2}$ below which no data can be acquired. This minimum value is the projection of the interferometer shortest baseline, and the corresponding angular extension in the real plane is called the `largest angular scale' (LAS).

The impossibility to measure the visibility function at small distances from the $uv$ plane origin has important consequences in the data imaging. The lack of data in this region of the $uv$ plane translates to a poor interferometer sensitivity to extended sources. There is no a net extension limit between sources that can be well imaged by interferometer and source that cannot be. We can safely state that only sources with angular dimension significantly smaller than the LAS can be reliably imaged by the interferometer, and that the more extended a source is the more poorly will be imaged.

The imaging artefacts produced by the missing low spatial frequencies prevent the total flux density recovery of the observed source. Therefore this issue, known as the `short-spacing problem' or the `flux-loss problem', prevents also a reliable reconstruction of the source spectral energy distribution.

\subsection{Single-dish observations and data reduction}
\label{sec:gbt_obs}
In order to fill in the gap in the central zone of the $uv$ plane and to cross-calibrate the single-dish and interferometer data, we needed to use a single-dish telescope whose diameter was much greater than the VLA minimum baseline, equal to $35\um{m}$ both in configuration C and D. For this reason we chose to observe with the GBT, which has an effective diameter of $100\um{m}$.

The observations were carried out in June 2011 at $1475\um{MHz}$ ($L$ band) and at $5100\um{MHz}$ ($C$ band), with both the receivers collocated in the Gregorian focus of the GBT. For the $L$ band the total bandwidth was $20\um{MHz}$, while for the $C$ band $80\um{MHz}$. As back-end we used the Digital Continuum Receiver.

The observing strategy was to map a sky region slightly wider than the VLA field of view. The GBT is equipped with a single-feed receiver but a map can be still obtained by letting the telescope move and scan the desired sky region. In particular we used the `on-the-fly mapping' technique: in this mode, the telescope is slewed within a rectangular area of the sky while it acquires the data and the map is built piling linear scans. Paying attention that, given the telescope beam, an appropriate sampling is made for each scan and that two adjacent linear scans are taken sufficiently close to each other, then the Nyquist sampling theorem guarantees that a reliable map of the sky can be obtained in the limits of instrumental errors. In order to reduce background and instrumental fluctuations, we set a slewing velocity high enough to execute scans of $0.3\um{s}$ per beam (whose width was chosen to be 3 pixels). This allowed us to reach the desired sensitivity of $5\um{mJy\ beam^{-1}}$ in $C$ band and to observe at the confusion level ($\sim20\um{mJy\ beam^{-1}}$) in $L$ band. The map dimensions are $25\times25\um{arcmin}^2$ for the $C$ band and $75\times75\um{arcmin}^2$ for the $L$ band. The scan separation is $50\um{arcsec}$ for $C$ band and $150\um{arcsec}$ for $L$ band, with the expected beam FWHM respectively $2.6\um{arcmin}$ and $9\um{arcmin}$.

\begin{figure}
\begin{center}
\includegraphics[width=\columnwidth]{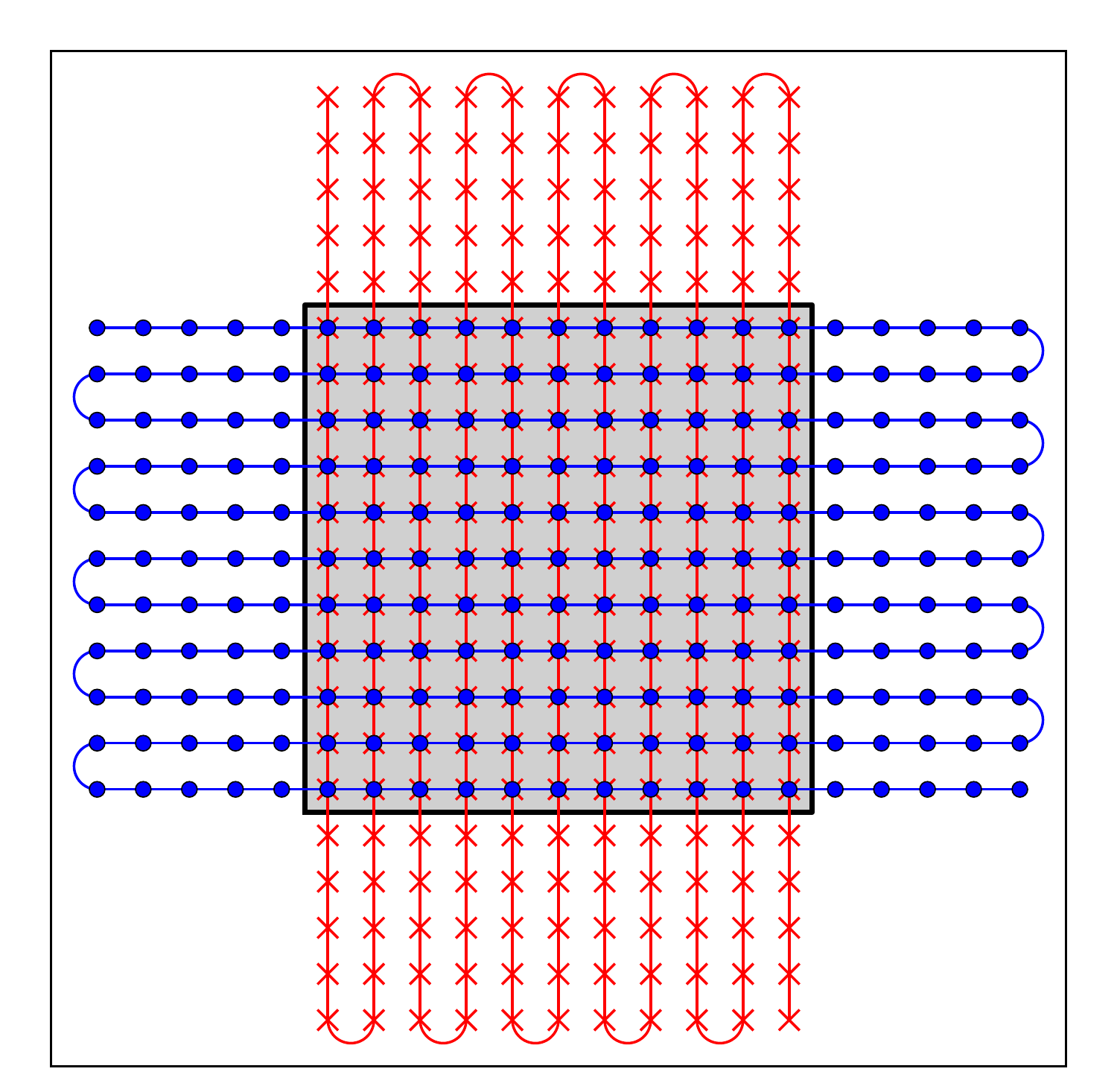}
\caption{A schematic illustration of the two intersecting scan series used in this work. The shadowed area is the sky region to map; first the telescope is moved along the horizontal path, with a data acquisition at each circle, then the piled scans are performed in the orthogonal direction, following the vertical path (crosses represent acquisition points).}
\label{fig:orthscan}
\end{center}
\end{figure}

The `piled-scans' mapping technique produces an important image artefact, with the resulting map affected by `stripes'. A method for stripe removal will be thoroughly discussed in Section \ref{sec:des} and in Appendix A1. This method requires that each field is mapped using two different and independent scan directions. We therefore mapped each field using two orthogonal complete scan series along right ascension and declination (see Figure \ref{fig:orthscan}). This procedure was repeated twice for each target source.

The entire data reduction process was performed using both routines provided by GBT staff and also by means of routines written by ourselves. First, the raw data from the telescope were converted to the SDFITS format, a particular, non-standard, FITS format explicitly designed to contain single-dish data in binary tables. On these files a flux calibration has been performed. Unavoidable telescope pointing errors did not allow the data acquisition positions to lie on the theoretical points planned and instructed to the telescope. We used interpolation via triangulation to regrid the data to the desired pattern (see Appendix A2). The calibrated and regridded data were eventually used to create a standard FITS image.

\section{Combining GBT and VLA maps}

\subsection{Destriping single-dish maps}
\label{sec:des}
The map construction method described in Section \ref{sec:gbt_obs} produces important artefacts in the final image. In particular, instrumental drifts during mapping time usually give rise to a different baselevel for each scan and the resulting image appears affected by `stripes', running in the same direction of the scans. The simplest way to take into account these baselevel drifts is to estimate their contribution in a region supposed to be free of emission. This can be obtained by letting the scans be more extended than the region to map and then using these extra-points, at both side of the map area, to linearly fit a baselevel that will be subtracted from each scan. In many circumstances, however, this method may not give a satisfactory result. \citet{Emerson1988} (EG88) pointed out three main limitations: first, there can be emission at the end of each scan; second, also the scan ends are affected by noise which can have great effects on the baseline determination; third, it is possible that a baseline cannot be adequately represented by a linear fit. In their paper, they discussed the general problem of the reduction of scanning noise in raster scanned data, with no limitations to the single-dish single-feed radio telescope case presented above. The solution proposed by EG88 was to filter out the spatial frequencies responsible for stripes by means of opportune weights to the data in the Fourier domain.

One of the most important problems when one maps a portion of the sky in the Galactic plane, with the resolution of the GBT, is that there are no large regions free of emission in the neighbourhood of any source. Therefore it is mandatory to use a destriping algorithm such as the one by EG88, since no reliable baselevel estimation at the ends of the scan can be made. However the strategy to create the maps with scans extended only few arcminutes with the telescope scanning quickly gives us two important advantages: the systematic errors along each scan can be considered constant and the small amount of pixels constituting the map allows to work more surgically in the Fourier plane. Therefore we modified the EG88 algorithm to better fit our problem. A detailed description of the adopted solution is reported in appendix A1.

The main advantage of this method, compared to the EG88 algorithm, is the preservation of the information from all spatial frequencies. In particular the central pixel in the Fourier plane maintains its value. This is very important because, as we said, its value is the sum of all pixel values and therefore if it were (almost) zero several negative-valued features would appear in the destriped image. Even if in the real image it accounts only for a flat base level (i.e. even a zero central pixel in the Fourier plane does not introduce further artefacts), it may play a role in the combination process. The preservation of the central pixel in the Fourier plane guarantees the map total flux density is conserved after the destriping procedure. However this could be not true for a single source. We tested this possible flux density alteration comparing sources from pre- and post-destriped maps (for details see Appendix A2). We found that the error introduced by the destriping process is negligible with respect to the other source of error discussed in this work (like calibration errors or background noise).

\subsection{Map combination}
\label{sec:comb}
As discussed, one of the goal of this work is to combine interferometric and single-dish maps. However, despite the presence of several works regarding the flux-loss problem and the combination of interferometric and single-dish data, only a few reliable implementations of the theoretical algorithms found in literature were available. Our choice was to use the new \textsc{casa} tool, \textsc{casafeather}, introduced in version 4.1.

The theory behind the combination we aimed to perform is rather simple. It is based on the fact that a single-dish telescope measures not only a single spatial frequency, but a whole range of continuous spatial frequencies up to a maximum one, which corresponds to the diameter of the dish, $D$. Hence, a single-dish telescope behaves as an interferometer with a continuous range of baselines, from zero to $D$. The Fourier transform of a single-dish image is a two-variable complex function that can be interpreted as the `intensity' of each bidimensional spatial frequency. On the other hand, an interferometer directly observes the source in the Fourier domain, measuring the visibility function, but it is not able to fill the entire Fourier plane, and in particular it will not cover the inner region of this plane up to a spatial frequency corresponding to the shortest baseline. This `hole' is usually filled during the imaging process through some kind of interpolation, which, however, cannot provide reliable values for the missing visibility (more precisely it should be treated as an extrapolation).

If the single-dish telescope diameter is greater than the shortest baseline of the interferometer, there would be an annular region in the Fourier plane inside which we have both interferometric and single-dish data for the source, i.e. the single-dish and the interferometer are sensitive to the emission related to some common spatial frequencies. The trick now is to use this overlapping region in the Fourier plane to cross-calibrate the data and make them as coincident as possible in this region. This operation is known as `feathering'. The \textsc{casafeather} tool allows the user to control some parameters involved in the feathering process. In particular we set the effective dish diameter to $80\um{m}$ and single-dish image scaling factor from 1 to $2.5$. We can then merge the two datasets to obtain a single visibility function defined from the zero baseline to the longest interferometer baseline. An inverse transformation of the visibility obtained in this way results in an image with the same resolution of the original interferometric image, but also with well determined low spatial frequencies. From this combined image a total flux density recovery is possible. It is hard to determine the reliability of this measurement. We found however that the flux densities should be corrected within 5 or 10 per cent, depending on the particular map (see Appendix A2).

The actual result of the combination varied appreciably for the different sources. For the two SNRs the combined maps show a significant improvement in terms of image quality as well as an increasing in total flux density. In particular the recovered extended emission is now clearly spatially coincident with the diffuse emission present in the MIPSGAL images. In the GBT images, both the two SNRs are characterised by a high total flux density at $5\um{GHz}$ ($>1\um{Jy}$ for both), are (among) the brightest sources in their neighbourhood and are clearly (at least at $5\um{GHz}$) detached by other extended sources. These three properties proved to be extremely important in the combination process, triggering between a useful combination and a disappointing result. All the other sources are less extended than the two SNRs. In these cases the GBT maps allowed us to exclude the presence of diffuse emission around these sources, reassuring that the VLA flux density estimates can be considered reliable.

\section{Discussion}
We discuss now on the results obtained for the two SNRs and the four PNe. We report a very brief summary of state-of-art knowledge for each one of them. We present then our maps and derive from them different physical parameters that complement or improve the values available in literature.

\subsection{SNR G011.2-00.3}
The SNR G011.2-00.3 is a well studied radio, IR and X-ray source. Very likely, it is the remnant of the historical SN 386 \citep{Reynolds1994}. The radio emission associated to this object has been first detected at $2.7$ and $5\um{GHz}$ by \citet{Altenhoff1970} and at $408\um{MHz}$ by \citet{Shaver1970}. It was proposed as a SNR by \citet{Dickel1972} from the lack of the $\mathrm{H}109\alpha$ line. The source was subsequently observed in a wide range of radio frequencies, from $30.9\um{MHz}$ to $32\um{GHz}$ (\citealt{Becker1975}; \citealt{Milne1979}; \citealt{Downes1984}; \citealt{Morsi1987}; \citealt{Kassim1988}). In particular the 32-GHz observations by \citet{Morsi1987} revealed a deviation from the spectral index calculated at the lower frequencies, interpreted as the evidence of a composite-type SNR \citep{Vasisht1996}. This implied also the presence of a central pulsar, discovered by \citet{Torii1997} with a period of $65\um{ms}$ and with a possible accretion disk \citep{Glushak2014}. The remnant is thought to have originated from a IIL/b SN \citep{Koo2007} and an early-B type progenitor \citep{Kothes2001}.

\begin{figure*}
\begin{center}
\includegraphics[height=7cm]{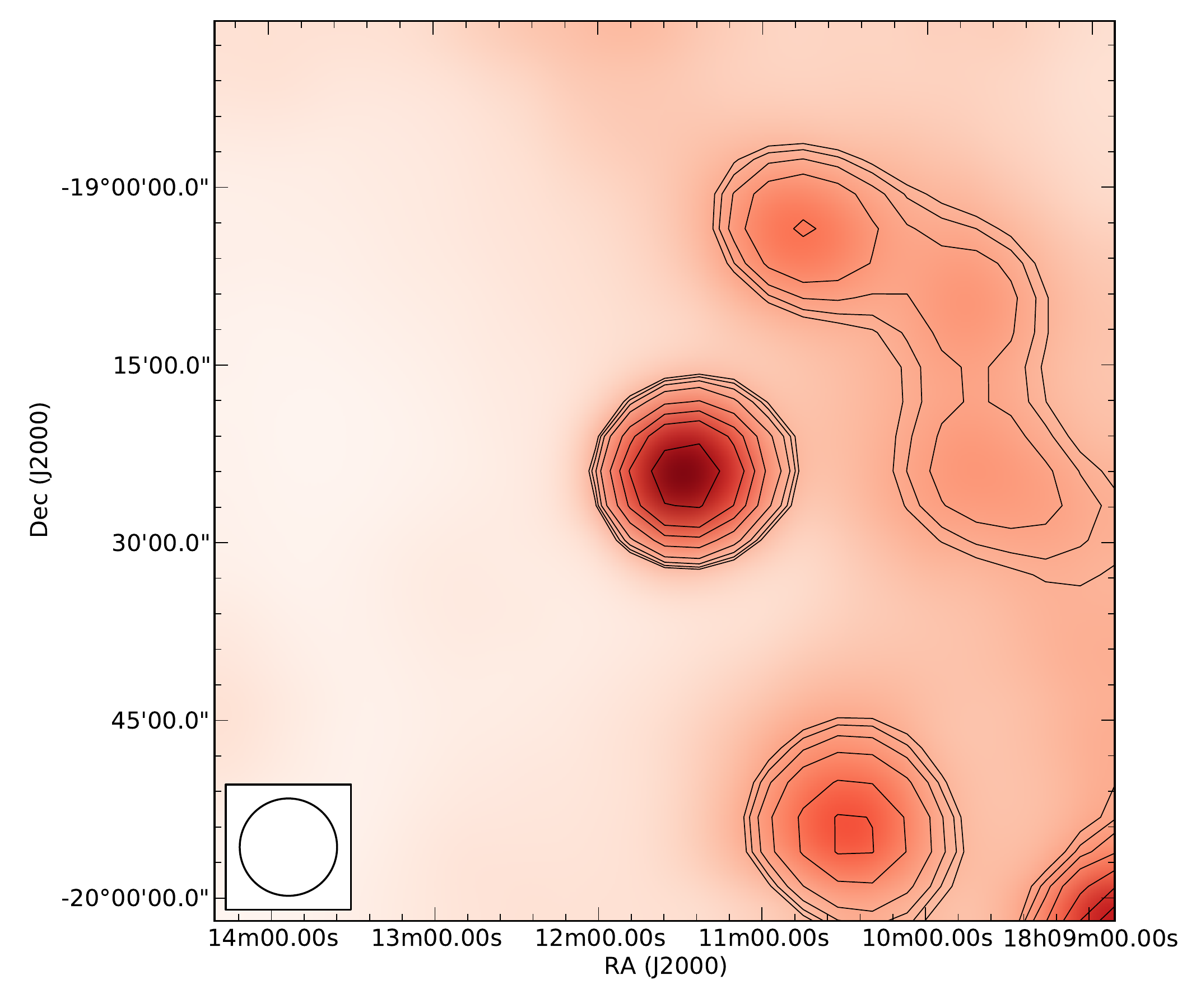}\hspace{1cm}
\includegraphics[height=7cm]{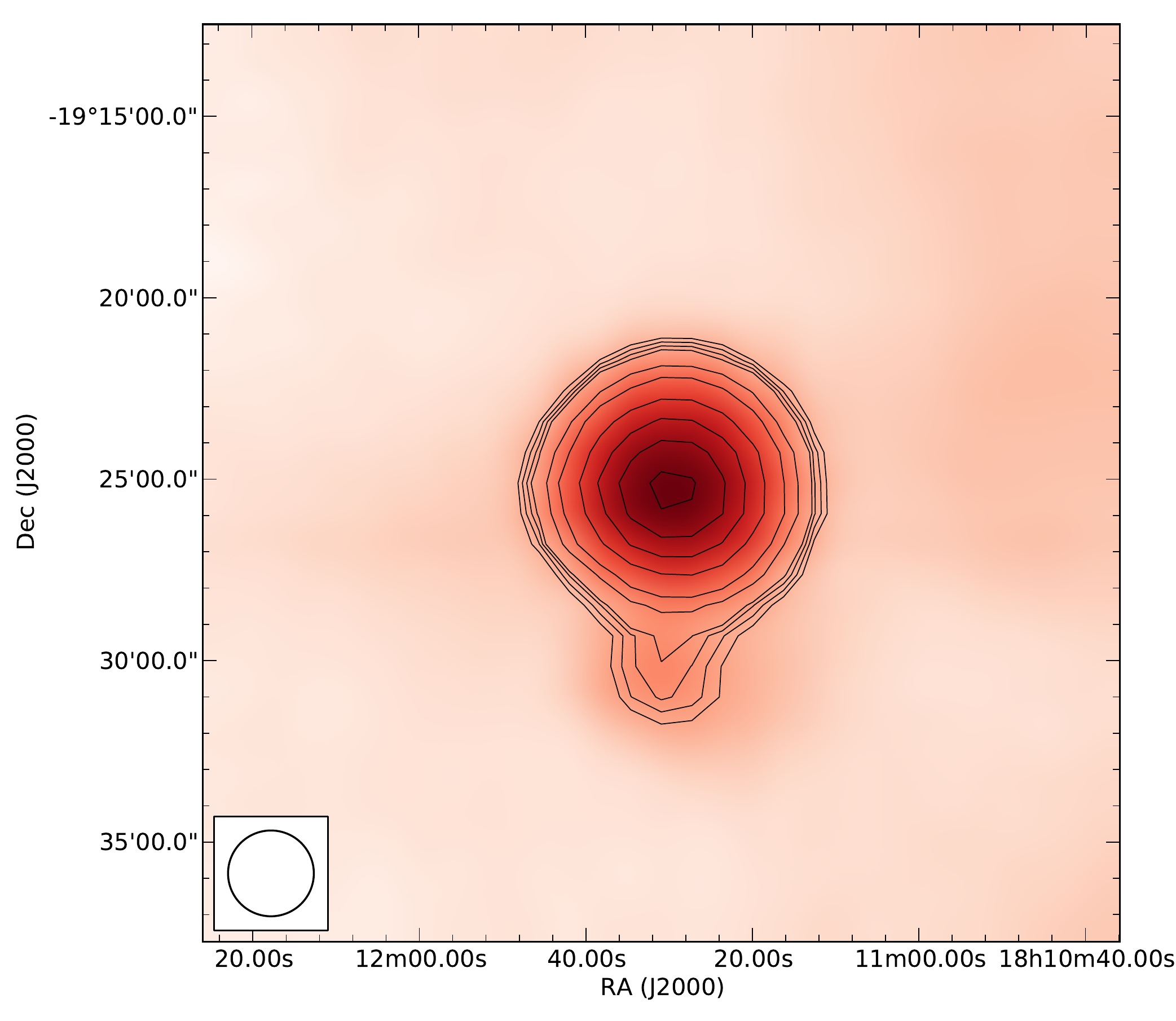}
\caption{GBT images of SNR G011.2-00.3 at $1.4\um{GHz}$ (left) and $5\um{GHz}$ (right). At $1.4\um{GHz}$ the source appears point-like and well distant from other sources. At $5\um{GHz}$ the source is resolved and another background source is clearly detected below the SNR. Please note that the scale of the two images is different.}
\label{fig:3910_gbt}
\end{center}
\end{figure*}

In Figure \ref{fig:3910_gbt} we report the GBT images obtained respectively at $1.4\um{GHz}$ and $5\um{GHz}$ after the destriping procedure described in Section \ref{sec:des}. At $1.4\um{GHz}$ the source is not resolved but appears quite far from nearby objects so that background estimation is not hard. At $5\um{GHz}$ the GBT beam is small enough to resolve the SNR (the other radio sources detected at $1.4\um{GHz}$ are not visible because they are out of boundary). However it is possible to notice a point source just below the SNR, too close to allow the huge beam at $1.4\um{GHz}$ to resolve the two distinct objects. In this case a flux density measurement at $1.4\um{GHz}$ only on GBT map could be overestimated.

\begin{figure}
\begin{center}
\includegraphics[width=\columnwidth]{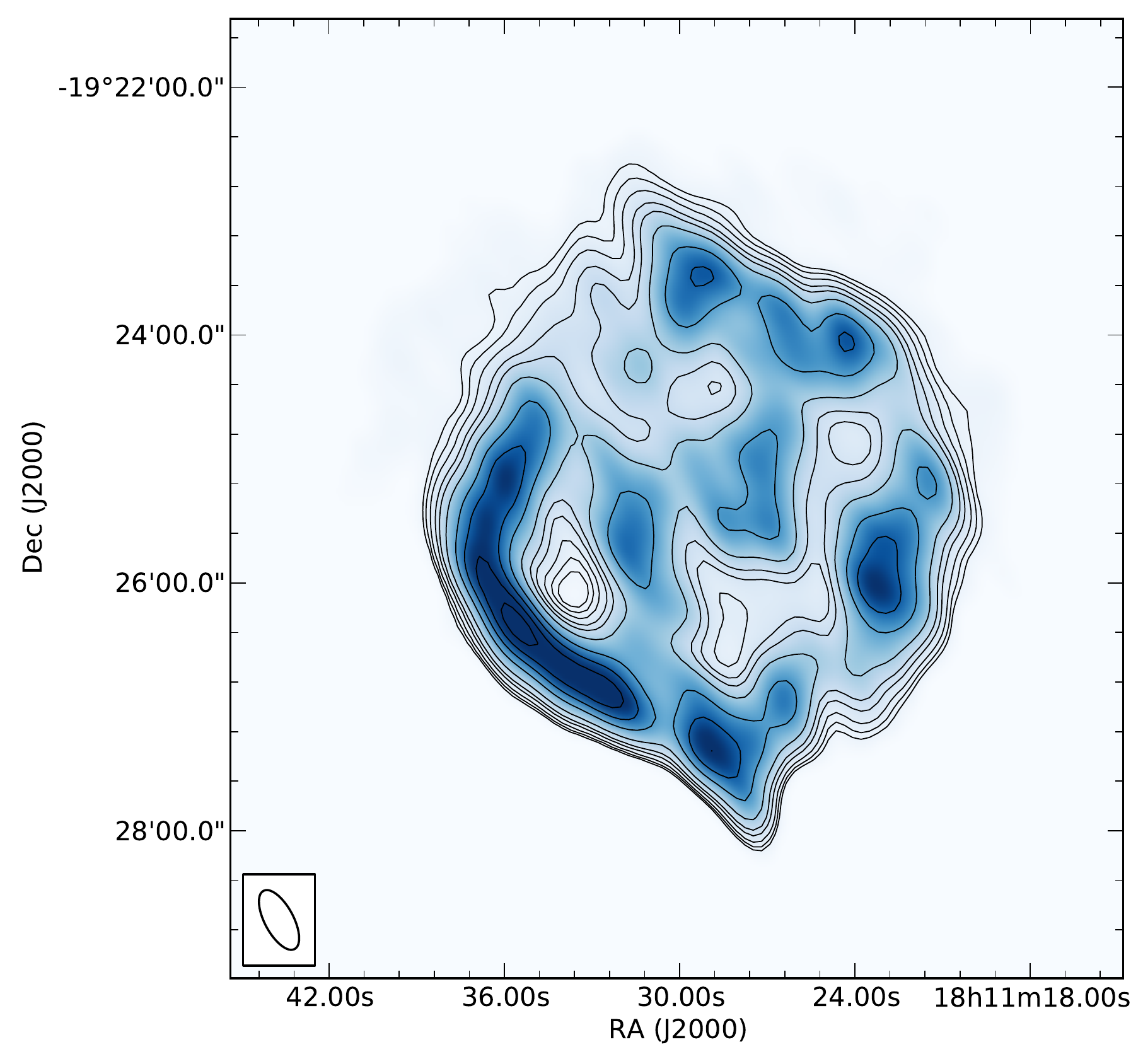}
\caption{High resolution images of SNR G011.2-00.3 at $5\um{GHz}$. The image is the result of feathering VLA (configuration D) and the GBT map.}
\label{fig:3910_gv}
\end{center}
\end{figure}

We produced a high resolution map at $5\um{GHz}$ combining the GBT data and the VLA map as described in Section \ref{sec:comb}. The resulting image is shown in Figure \ref{fig:3910_gv}. The composite-type nature of the SNR is evident, with a prominent shell and a perfectly recognisable pulsar wind nebula (PWN) at its center. We calculated a flux density of $8.2\pm0.8\um{Jy}$, in agreement with previous determination \citep{Downes1984}. In contrast, the flux density determined only on the VLA data was about $3\um{Jy}$, showing an important flux loss. It is extremely important to notice that our flux density determination derives from the combination of single-dish and interferometer data, and the fact that its value agrees with single-dish estimates from literature strongly corroborates the entire data reduction process. This excellent reliability is fundamental for more quantitative analysis, as the one presented in the next section for the other SNR.

\subsection{SNR G027.4+00.0}
The radio emission associated to SNR G027.4+00.0, also known as KES 73, is located in a particularly crowded region. At radio wavelengths, the entire region was first detected as a single discrete source at $408\um{MHz}$ \citep{Large1961}. Higher resolution images at $5\um{GHz}$ resolved the huge radio emission region in at least four discrete components. One of them presented a clearly non-thermal emission and was proposed as a SNR (\citealt{Milne1969}; \citealt{Angerhofer1977}).

\citet{Kriss1985} gave the first evidence that the SNR harboured a compact source, by means of X-ray observations. This compact source, designed 1E 1841-045, is an anomalous X-ray pulsar with a period of $11.8\um{s}$ \citep{Vasisht1997}. It is characterised by an extremely intense magnetic field \citep{Gotthelf1999} and shows soft $\gamma$-ray repeater like bursts (\citealt{Kumar2010}; \citealt{Lin2011}). It is now an established magnetar \citep{Kumar2014}.

Studies on H \textsc{i} absorption line toward the source pose the SNR at a distance of about $7.5\um{kpc}$ (\citealt{Sanbonmatsu1992}; \citealt{Tian2008}). The nebula originated $\lesssim\!2000\um{yr}$ ago from a core collapsing SN (type II or Ib) and a progenitor mass $\gtrsim\!20M_\odot$ (\citealt{Gotthelf1997}; \citealt{Tian2008}; \citealt{Kumar2014}).

\begin{figure*}
\begin{center}
\includegraphics[height=7cm]{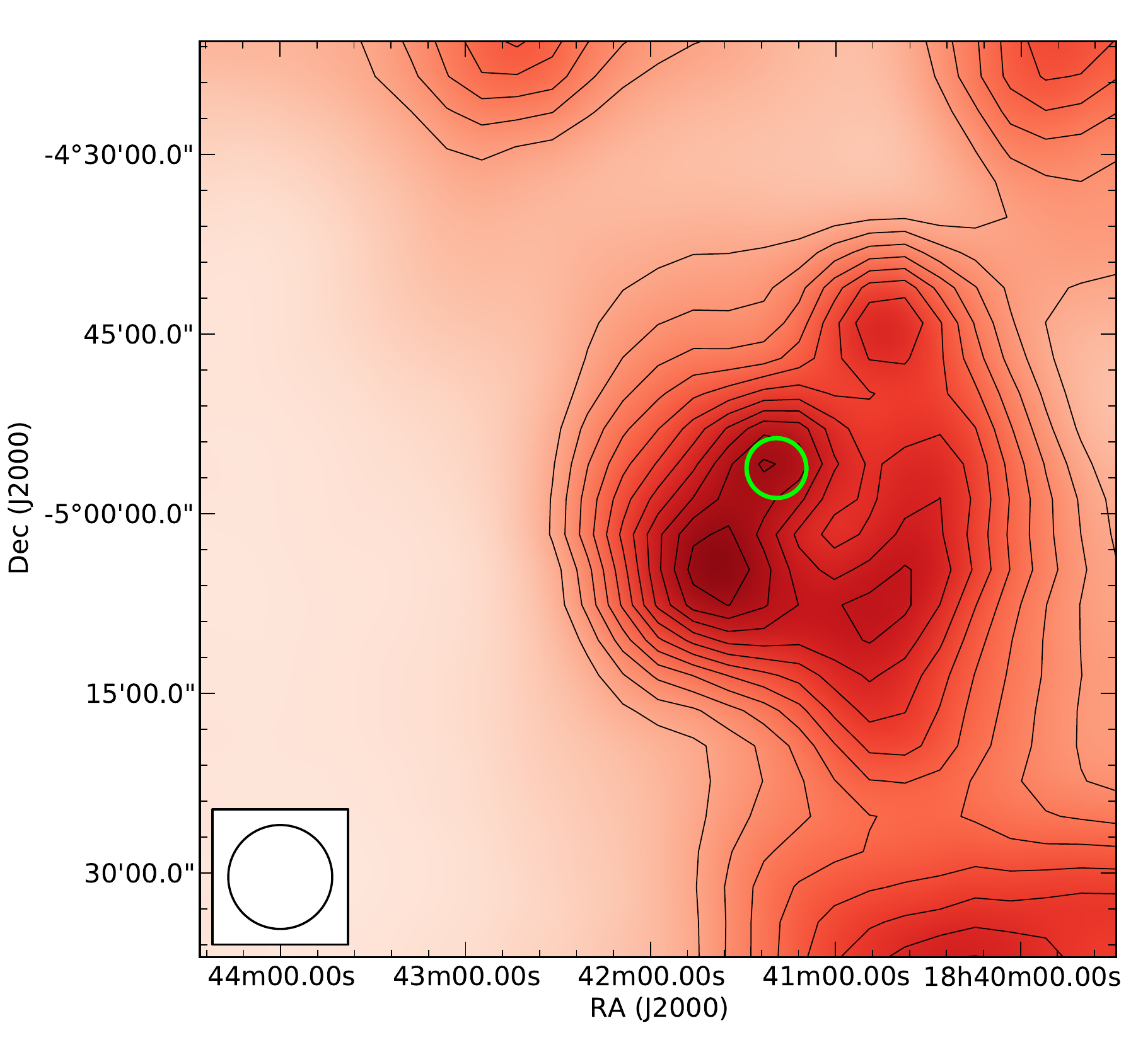}\hspace{1cm}
\includegraphics[height=7.2cm]{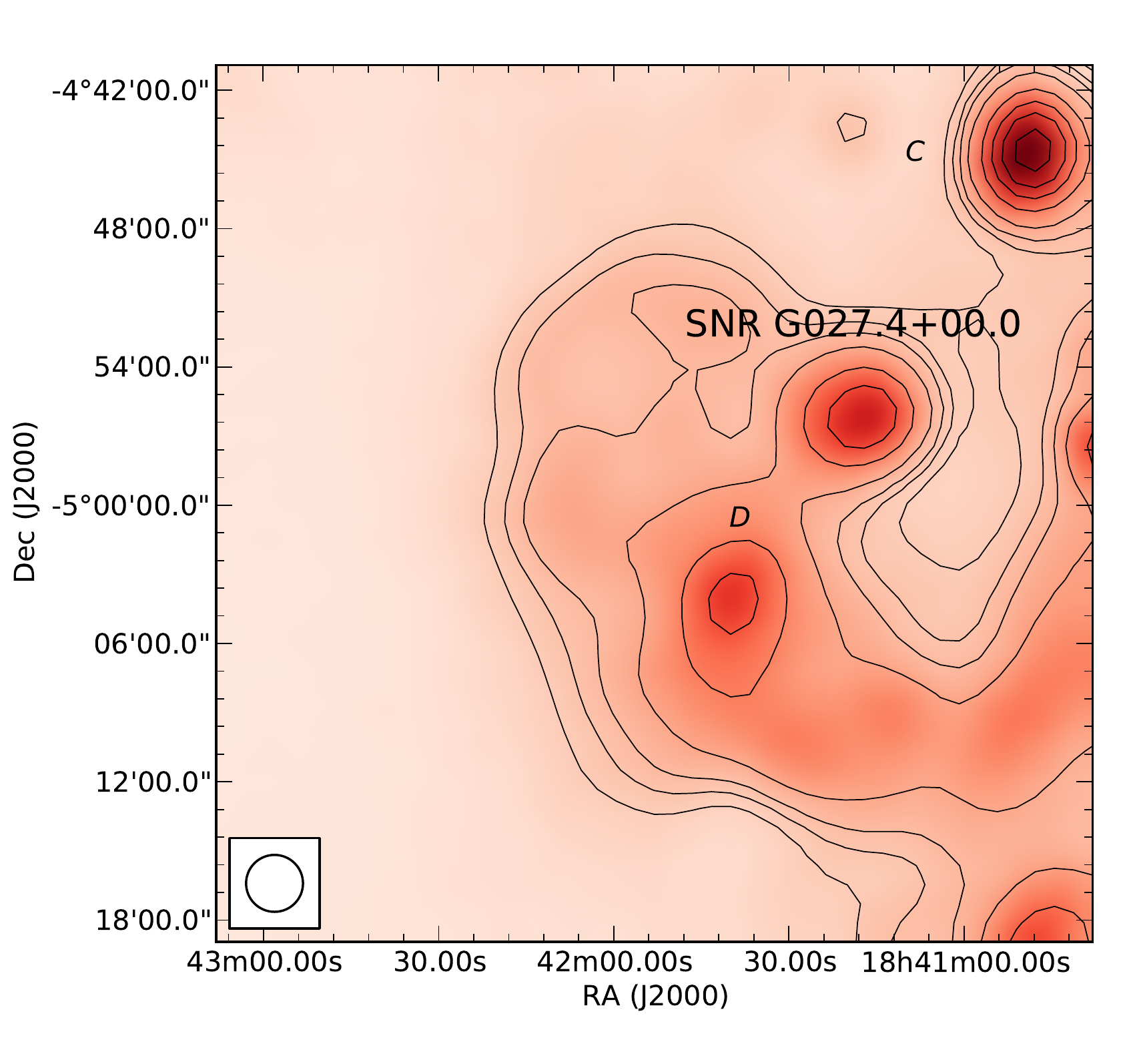}
\caption{GBT images of SNR G027.4+00.0 at $1.4\um{GHz}$ (left) and $5\um{GHz}$ (right). At $1.4\um{GHz}$ our map cover the entire radio emitting region described by \citet{Large1961}, and the SNR is highlighted with a green circle. At $5\um{GHz}$ we were able to clearly separate the SNR from the remainder of the other radio sources, as well as resolve the source itself. Labels $C$ and $D$ are as in \citet{Angerhofer1977}. Please note that the scale of the two images is different.}
\label{fig:3310_gbt}
\end{center}
\end{figure*}

In Figure \ref{fig:3310_gbt} we report the GBT images we obtained respectively at $1.4\um{GHz}$ and $5\um{GHz}$. It is possible to notice how the radio emission from the surrounding sources discussed above is clearly detected. At $5\um{GHz}$ the GBT is able to easily resolve the SNR from the remainder of radio sources while at $1.4\um{GHz}$ the SNR is not totally detached.

\begin{figure*}
\begin{center}
\includegraphics[height=6.7cm]{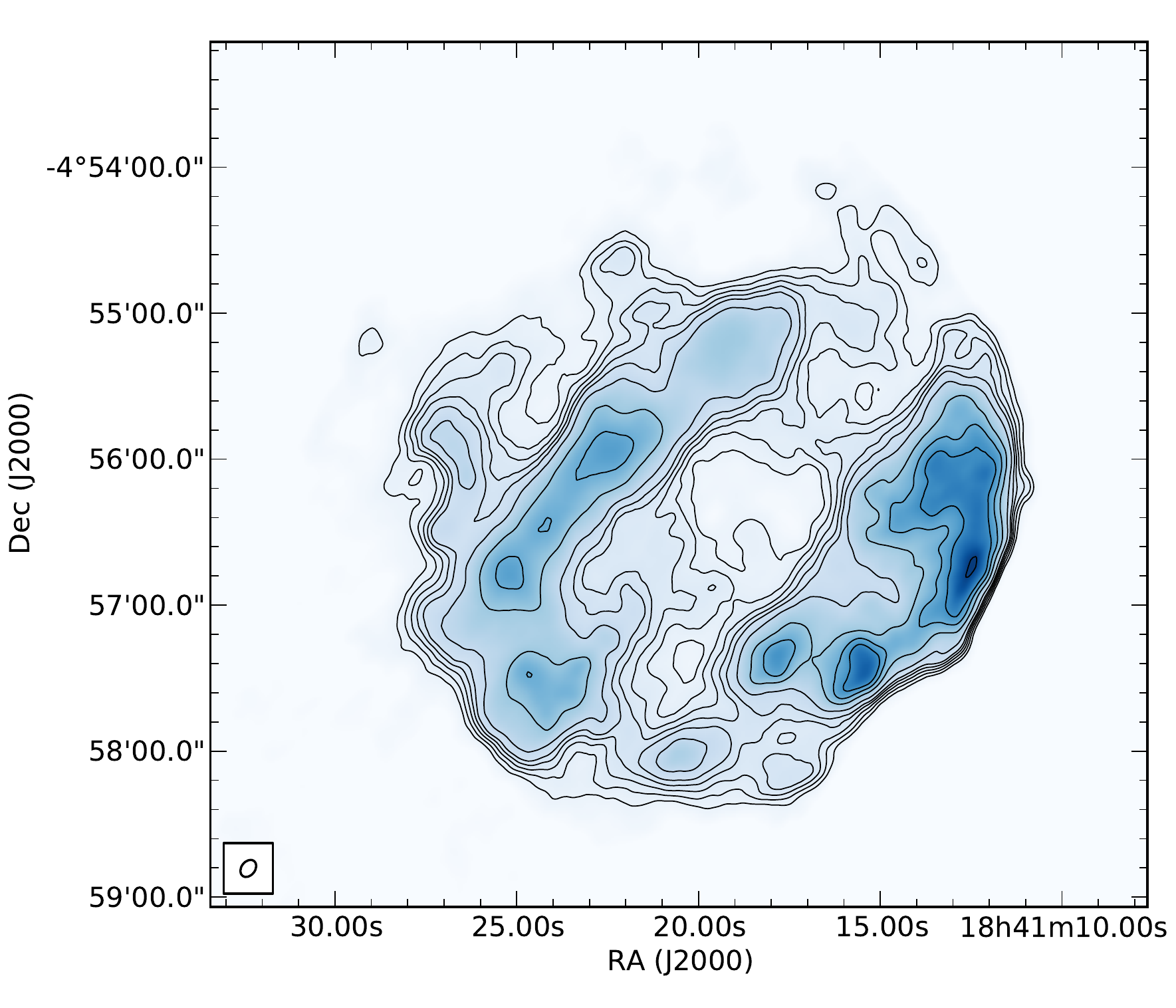}\hspace{1cm}
\includegraphics[height=6.6cm]{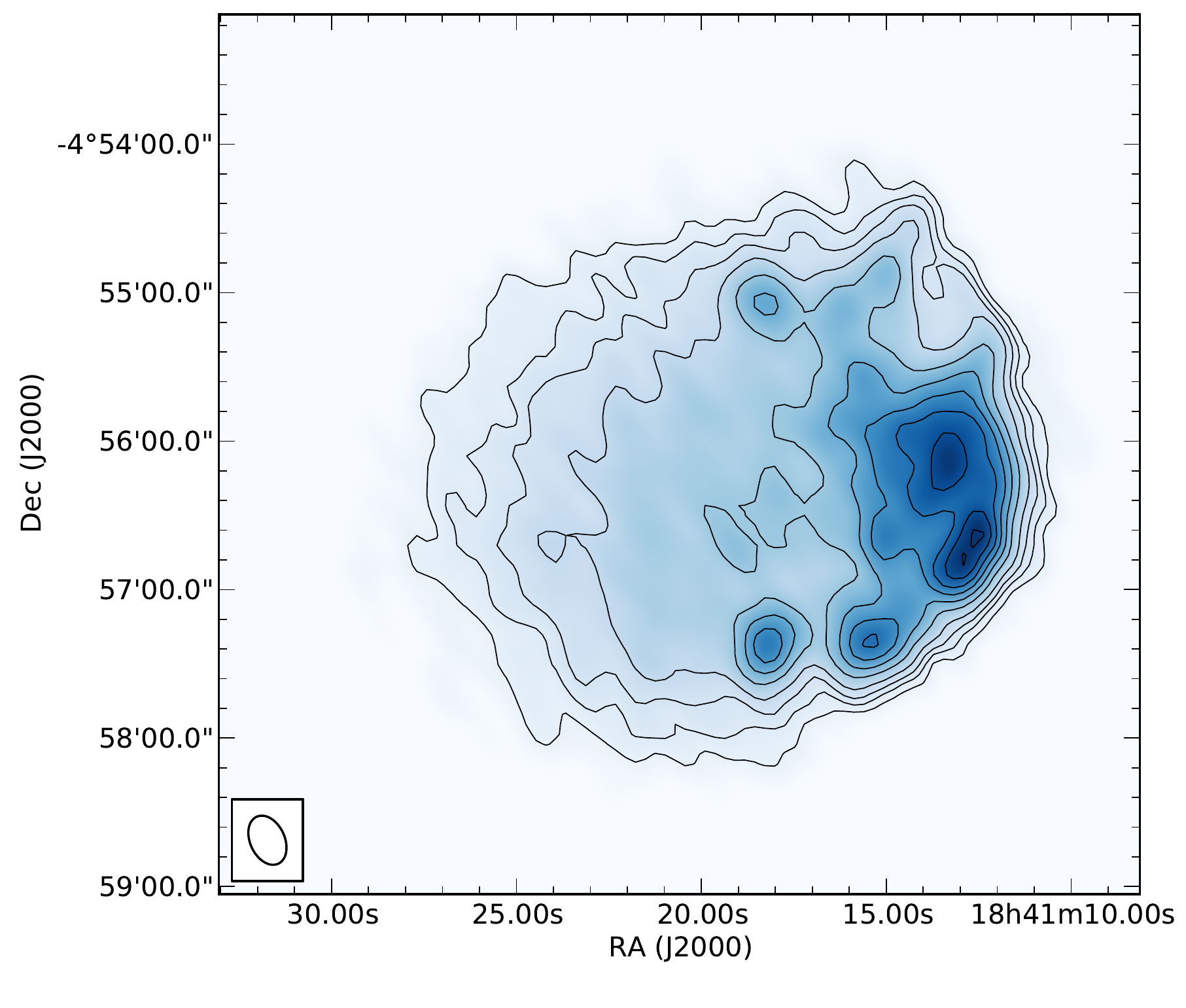}
\caption{High resolution images of SNR G027.4+00.0 at $1.4\um{GHz}$ (left) and $5\um{GHz}$ (right). The $1.4\um{GHz}$ image was obtained combining GBT and VLA B- and C-configuration data. The $5\um{GHz}$ image is the result of feathering VLA (configuration D) and the GBT map.}
\label{fig:3310_highres}
\end{center}
\end{figure*}

A high resolution map at $1.4\um{GHz}$ has been created using VLA data from both our observations (configuration C; Section \ref{sec:vla_obs}) and archive data (configuration B), and combining them via \textsc{casafeather}. Though the source is less extended than the VLA LAS at this frequency we further combined this high resolution image with the GBT map. The result is shown in Figure \ref{fig:3310_highres} (left). The calculation of the total flux density of the SNR proved hard because of the peculiar background. We measured a value of $5.5\pm0.8\um{Jy}$, compatible with the value in literature \citep{Koo1991}. There is no evidence of radio emission from the pulsar, with a flux density upper limit of $0.33\um{mJy}$ at $3\sigma$ level (previous estimates gave an upper limit of $0.60\um{mJy}$ at this frequency, see \citealt{Kumar2014}).

At $5\um{GHz}$ we did not find useful archive observations and we created a high resolution map combining the GBT map and the VLA data (configuration D; Section \ref{sec:vla_obs}). The combined image is shown in Figure \ref{fig:3310_highres} (right). In this case the source was more extended than the VLA LAS and the availability of single-dish data proved crucial to recover the total flux density. For this frequency we calculated a flux density of $2.3\pm0.4\um{Jy}$, in accordance with previous estimates (\citealt{Angerhofer1977}; \citealt{Caswell1982}). For comparison, the flux density computed only on the VLA map was $0.51\pm0.01\um{Jy}$, and therefore, in this case, we can see a flux-loss of at least a factor 4.

\begin{figure}
\begin{center}
\includegraphics[width=\columnwidth]{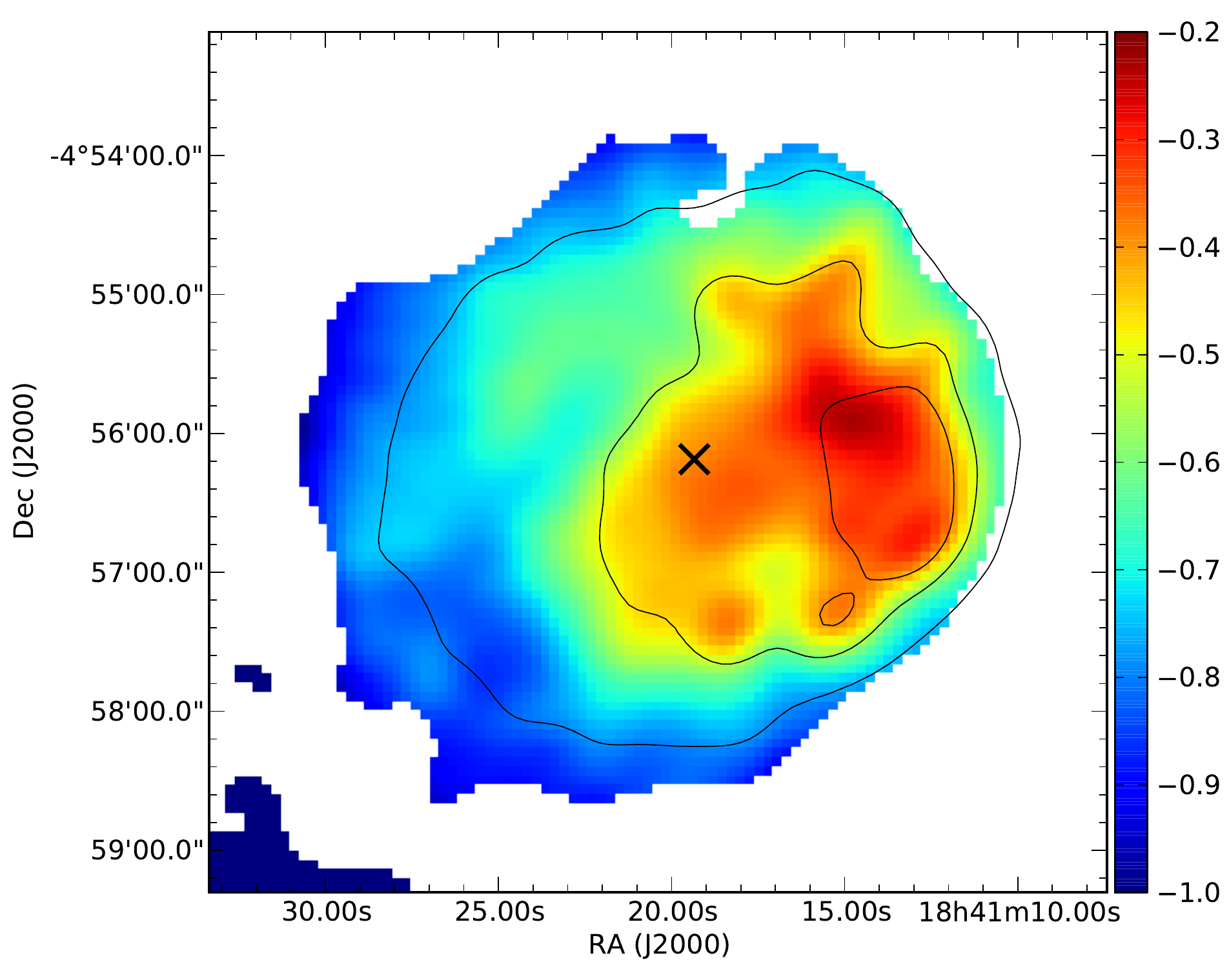}
\caption{Spectral index map of SNR G027.4+00.0. The super-imposed contours refer to the 5-GHz map convolved to a circular 25-arcsec beam (see text). The pulsar is indicated with a black cross.}
\label{fig:spix}
\end{center}
\end{figure}

\begin{figure*}
\begin{center}
\includegraphics[height=6.7cm]{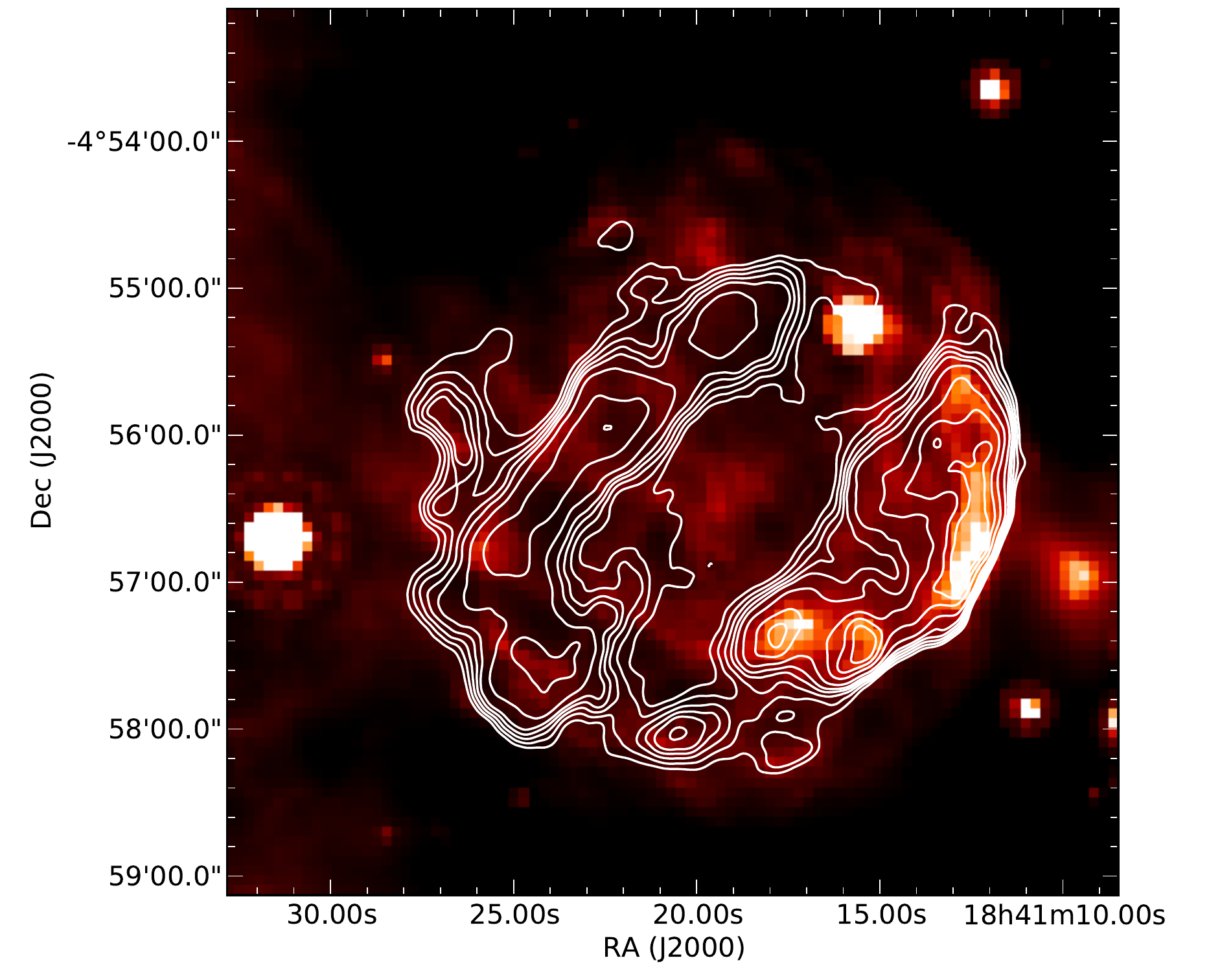}\hspace{1cm}
\includegraphics[height=6.7cm]{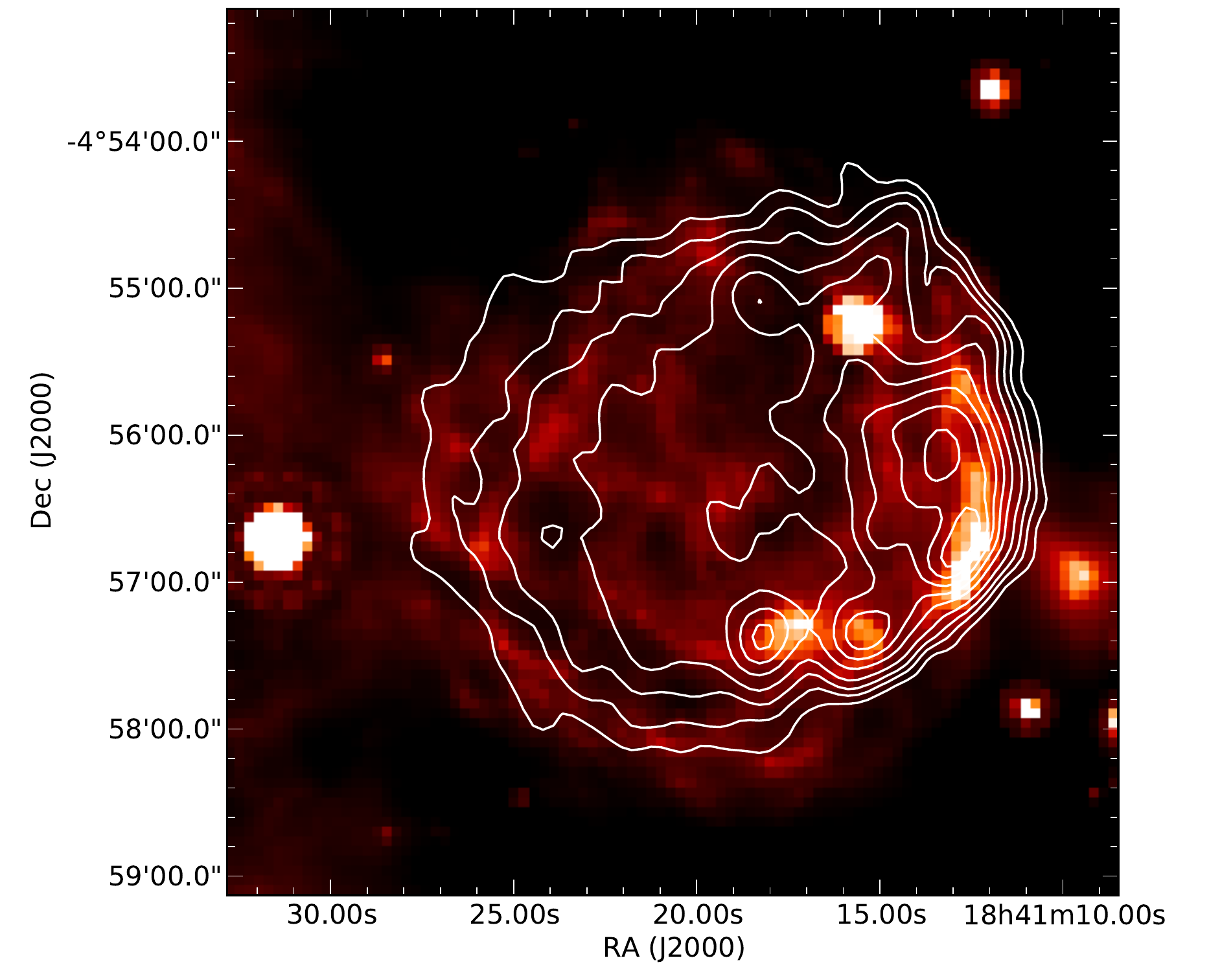}
\caption{Superposition of the MIPSGAL 24-$\umu$m image of SNR G027.4+00.0 and radio contours at $1.4\um{GHz}$ (left) and $5\um{GHz}$ (right). The images at $24\mic{m}$ and at $1.4\um{GHz}$ have approximately the same resolution. We can notice that the western region appears bright both in IR and in radio. The north-west region and the central part of the SNR are those with the flattest spectrum. In these two regions the IR emission seems co-spatial with the 5-GHz one but the poorer resolution of the latter does not permit definitive conclusions. The bright IR source in the north-west region of the SNR is the unrelated star IRAS 18385-0458.}
\label{fig:sup}
\end{center}
\end{figure*}

Given the flux densities computed above we can derive a global spectral index $\alpha=-0.70\pm0.18$, in perfect agreement with previous estimates in other spectral region (the turnover is around $100\um{MHz}$; \citealt{Caswell1982}; \citealt{Kassim1989}). Starting from the two images presented in Figure \ref{fig:3310_highres} we were also able to build a spectral index map. In particular we convolved both the images with an opportune bidimensional Gaussian in order to obtain for both a circular beam of $25\um{arcsec}$. The convolved images were subsequently regridded to have a pixel-by-pixel correspondence and the spectral index map was created. The result is shown in Figure \ref{fig:spix}. Not only a prominent spectral index variation is evident but it is also possible to notice a flatter index toward the pulsar with respect to the global value, as well as an important asymmetry, with the western region flatter than the eastern one. A spectral index of about $-0.3$ around the pulsar is usually associated to the presence of a PWN \citep{Gaensler2006}. However this should not be the case since this SNR is classified as a pure shell-type remnant and, furthermore, PWNe surrounding a magnetar have never been detected. The western region appears the brightest one not only in radio but also in IR and X-ray and it is characterised by a higher column density with respect to the other regions \citep{Kumar2014}. Being the compact star a magnetar, it is possible to hypothesise that its strong magnetic field has an influence on the western region, increasing the remnant intensity. Moreover, spatial inhomogeneities in the magnetic field intensity lead to variations of the emission turnover. Overlapping synchrotron contributions with different turnover frequencies could finally result in a flatter spectral index. Alternatively, \citet{Giacani2011} discussed a similar picture for the shell-type SNR G344.7-00.1, where a flat spectral index around $-0.3$ is measured toward the central regions. They notice that in that region there is an important correspondence between the radio and the 24-$\umu$m emission. They conclude that the flatter spectral index derives from the SN shock impacting a dense molecular cloud, resulting in a radiatively energy loss. In Figure \ref{fig:sup} we show a superposition of MIPSGAL 24-$\umu$m image of SNR G027.4+00.0 with our radio maps. We can notice that in the western region the IR and the 1.4-GHz emission are coincident. A morphological comparison between 24-$\umu$m and 5-GHz image is more difficult because of the poor resolution of the radio map. The 5-GHz emission seems to trace well the 24-$\umu$m image in the north-west part of the SNR (which is the region with the flattest spectrum) and partially in its centre. A satisfactory morphological analysis would require that also 5-GHz map had about the same resolution of the IR image. Though the correlation between IR and radio for our SNR appears less stringent than for SNR G344.7-00.1 we cannot rule out the hypothesis of shock impacting dense molecular cloud as a cause of spectral flattening. Finally another limb-flattened spectral index behaviour is reported by \citet{Bhatnagar2011} for the composite-type SNR G016.7+00.1.

\subsection{Planetary nebulae}
One of the main conclusions of Paper I was that the great majority of the MIPSGAL bubbles showing radio emission could be classified as PNe. In that sense, the radio study of these objects had proved an extremely powerful instrument for finding undiscovered PNe (currently only about 10 percent of the expected galactic PNe have been found). However the sole IR and radio images were insufficient to determine many physical characteristics of these PNe and we could not go further than the mere, sometimes tentative, classification.

In this section we present the PNe in our sample. We derive some physical properties starting from the radio maps, given their electron temperature (see, for example, \citealt{Wilson2012}). For each PN, we consider the 5-GHz map and we assume that at this frequency the nebula is an optically-thin free-free emitter, as derived from their spectral index (see Table \ref{tab:PNe}). Under this assumption the nebula brightness $B$ is given by
\begin{equation}
B=B_\mathrm{bb}(T_e)\tau_\nu,
\end{equation}
where $T_e$ is the electron temperature, $B_\mathrm{bb}(T_e)$ is the brightness of a black body at a temperature $T_e$ and $\tau_\nu$ is the optical depth at the frequency $\nu$. If $T_e$ is known, from the brightness map it is possible to derive an optical-depth map. The optical depth is related to the emission measure $\mathrm{EM}$ through the relation
\begin{equation}
\tau_\nu=8.235\times10^{-2}T_e^{-1.35}\left(\frac{\nu}{\mathrm{GHz}}\right)^{-2.1}\left(\frac{\mathrm{EM}}{\mathrm{pc\cdot cm^{-6}}}\right).
\end{equation}
If also the distance is known it is possible to estimate the dimension of the nebula. In this case the electron density $n_e$ can be derived from the relation
\begin{equation}
\mathrm{EM}=\int_0^s\!\!\!n_e^2ds,
\end{equation}
where $s$ is the line-of-sight coordinate. Assuming an opportune geometry for each PN and that the nebular gas is completely ionized hydrogen, it is finally possible to derive the PN ionized mass as a volume integral:
\begin{equation}
M_\mathrm{ion}\approx\int_Vn_em_pdV,
\label{eq:mion}
\end{equation}
where $m_p$ is the proton mass.

From the equation (\ref{eq:mion}), and considering the previous ones, it is possible to show that
\begin{equation}
M_\mathrm{ion}\propto T_e^{0.35}D^{5/2},
\end{equation}
where $D$ is the distance. Therefore, the ionized mass shows only a weak dependence on $T_e$, while it is very sensitive to the nebula distance. The distance values from literature usually suffer of important uncertainties. Given the nebula flux density at $5\um{GHz}$ and its mean angular radius, we used the formula derived by \citet{VandeSteene1995}, which holds only for optically thin sources, to calculate the PN statistical distance. We adopted this value as the best distance estimate, except for the last PN (see Section \ref{sec:3706}). The less critical dependence on $T_e$ allows us to safely assume a typical value of $10^4\um{K}$ whenever no previous estimates were available.

In the Table \ref{tab:PNe} we report the mean electron density, the distance and the ionized mass that we derived for the four PNe. A brief discussion on each one of them follows.

\begin{table}
\begin{center}
\begin{tabular}{lccccc}\hline\hline
Source & $\langle n_e\rangle$ & $D$ & $T_e$ & $M_\mathrm{ion}$ & $\alpha$\\
& ($\mathrm{cm}^{-3}$) & (kpc) & ($10^3\um{K}$) & ($M_\odot$)\\\hline
PN G029.0+00.4 & 39.2 & 1.6 & $\phantom{1}7.5$ & 0.1 & $-0.09$\\ 
NGC 6842 & 15.6 & 2.2 & 13.2 & 0.1\\ 
PN G040.3-00.4 & 29.3 & 2.1 & \textit{10.0} & 0.2 & $-0.06$\\ 
PN G031.9-00.3 & 14.3 & 3.2 & \textit{10.0} & 0.1 & $\phantom{-}0.49$\\ 
\hline
\end{tabular}
\label{tab:PNe}
\end{center}
\caption{Physical parameters for the four PNe. The first column is the mean electron density as derived from our 5-GHz map. In the second we report the distance value used to compute the ionized mass. The third column lists the electron temperature as reported in literature for the first two PNe and the assumed value of $10^4\um{K}$ for the other two (see text). In the fourth column we report the ionized mass while in the last one the radio spectral index.}
\end{table}

\subsubsection{PN G029.0+00.4} 
Also known as Abell 48, PN G029.0+00.4 was first discovered thanks to the National Geographic Society--Palomar Observatory Sky Survey and recognised as PN after its morphology (\citealt{Abell1955}; \citealt{Abell1966}). Optical spectroscopy of the central star by \citet{Wachter2010} seemed to dispute its nature of PN in favour of a WR star with a surrounding nebula. However, recently, \citet{Todt2013} and, independently, \citet{Frew2014} (F14 hereafter) conducted extremely deep spectroscopic studies both on the central source and on the nebula. The two groups came to the same conclusion that Abell 48 is indeed a PN harbouring an extremely rare [WN5] or [WN4-5] star. The distance of PN G029.0+00.4 has been estimated between 1.6 and $1.9\um{kpc}$ (\citealt{Cahn1971}; \citealt{Ortiz2013}; \citealt{Todt2013}; F14).

This object was observed in the IR by \textit{Spitzer} (both IRAC and MIPS), \textit{WISE} and \textit{AKARI} (\citealt{Phillips2008}; \citealt{Phillips2011}). The mid-IR infrared colour, in particular, was used by F14 as one of the indicator of the PN nature. The first radio detection was made by \citet{Cahn1974} at $11.1\um{cm}$ and then in the NVSS at $1.4\um{GHz}$ \citep{Condon1998b}.

\begin{figure}
\begin{center}
\includegraphics[width=\columnwidth]{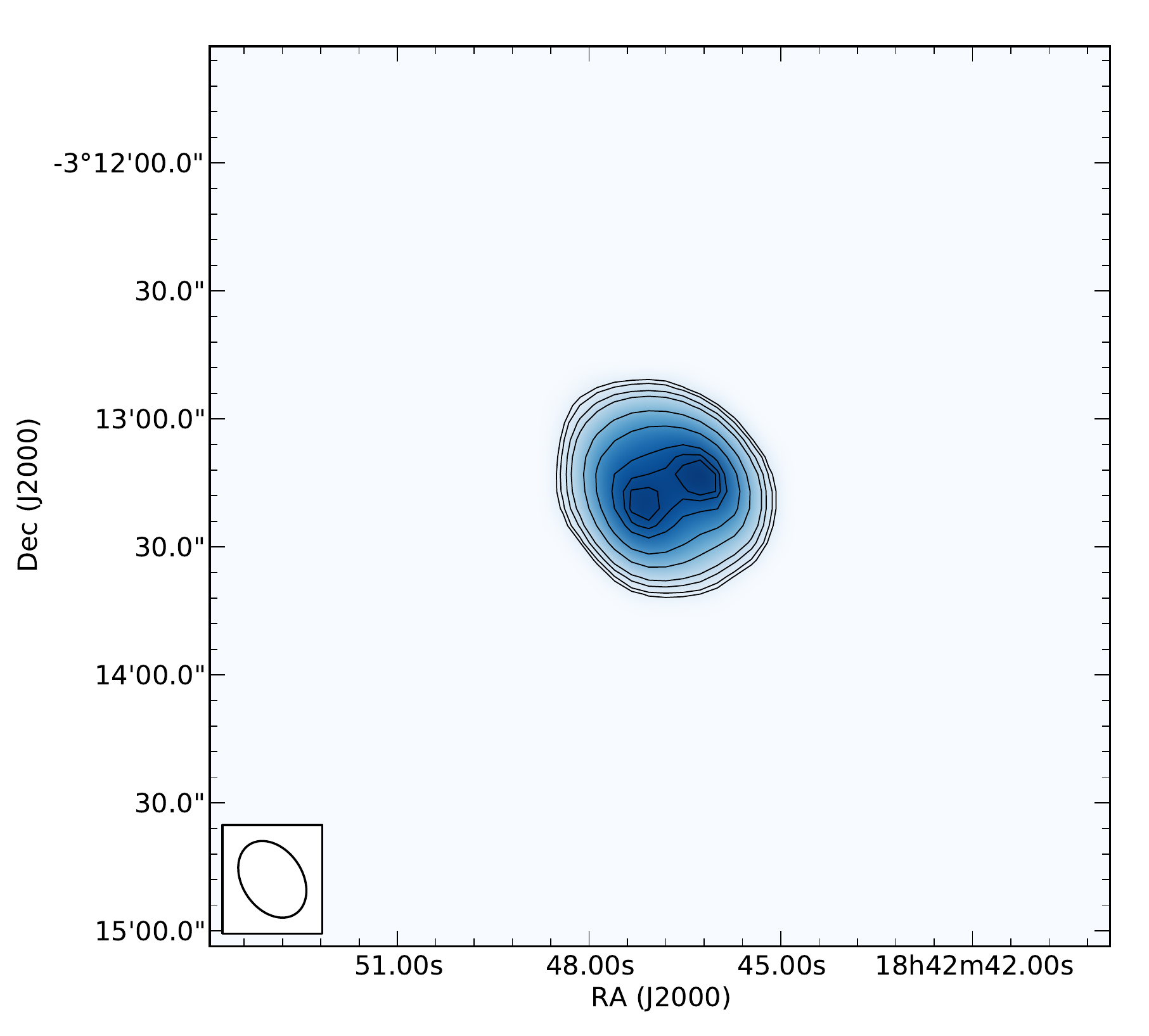}
\caption{VLA image of PN G029.0+00.4 at $5\um{GHz}$.}
\label{fig:3328_highres}
\end{center}
\end{figure}

Our VLA map at $5\um{GHz}$ is presented in Figure \ref{fig:3328_highres}. The source extension ($\sim\!1\um{arcmin}$) is well below the LAS of the VLA in configuration D, therefore a combination with the GBT data proved unnecessary. Indeed a test combination was performed and, as expect, no significant variations were revealed. The resolution achieved in Figure \ref{fig:3328_highres} ($19.5\times14\um{arcsec}^2$) is worse than the resolution of NVSS archive image at $1.4\um{GHz}$, however the bipolar structure of the PN is still recognisable. For this source, we estimated a flux density at $5\um{GHz}$ $S_C=142.4\pm4.3\um{mJy}$. Using the NVSS flux density value at $1.4\um{GHz}$ $S_L=159\pm15\um{mJy}$, we derived a spectral index $\alpha=-0.09\pm0.08$, in accordance with a optically-thin free-free emission. From the 5-GHz flux density we derived a statistical distance of $1.6\pm0.2\um{kpc}$.

F14 were able to estimate the ionized mass of the nebula exploiting the integrated H$\alpha$ flux density. They found $M_\mathrm{ion}\!\sim\!0.3\sqrt{\epsilon}M_\odot$, where $\epsilon$ is the nebula filling factor. We used the map at $5\um{GHz}$ to provide a different approach to the same problem. Assuming an electron temperature of $7500\um{K}$ (F14) and a depth equal to its mean diameter (this assumption holds also for the following PNe), we found $M_\mathrm{ion}\!\sim\!0.12M_\odot$. This value is in agreement with the one from F14 and is compatible with the typical ionized mass values for PNe \citep{Frew2010}.

\subsubsection{NGC 6842} 
First listed in the New General Catalogue \citep{Dreyer1888}, NGC 6842 was classified as PN by \citet{Curtis1919} because of its morphology. It appears as a ring nebula with a diameter of $\sim\!50\um{arcsec}$ \citep{Hromov1968}. Despite the fact that several studies have dealt with its distance determination, to date there is no a strong agreements among the various estimates, with values ranging from 1.25 to $2.9\um{kpc}$ (\citealt{Odell1962}; \citealt{Cahn1971}; \citealt{Zhang1995}; \citealt{Gorny1997}; \citealt{Tajitsu1998}; \citealt{Phillips2004}; \citealt{Giammanco2011}).

The source has been observed at radio frequencies from $327\um{MHz}$ to $5\um{GHz}$ (\citealt{Taylor1983}; \citealt{Gregory1986}; \citealt{Taylor1996}; \citealt{Condon1998b}), and its nature of thermal emitter has been proven \citep{Taylor1996}.

\begin{figure}
\begin{center}
\includegraphics[width=\columnwidth]{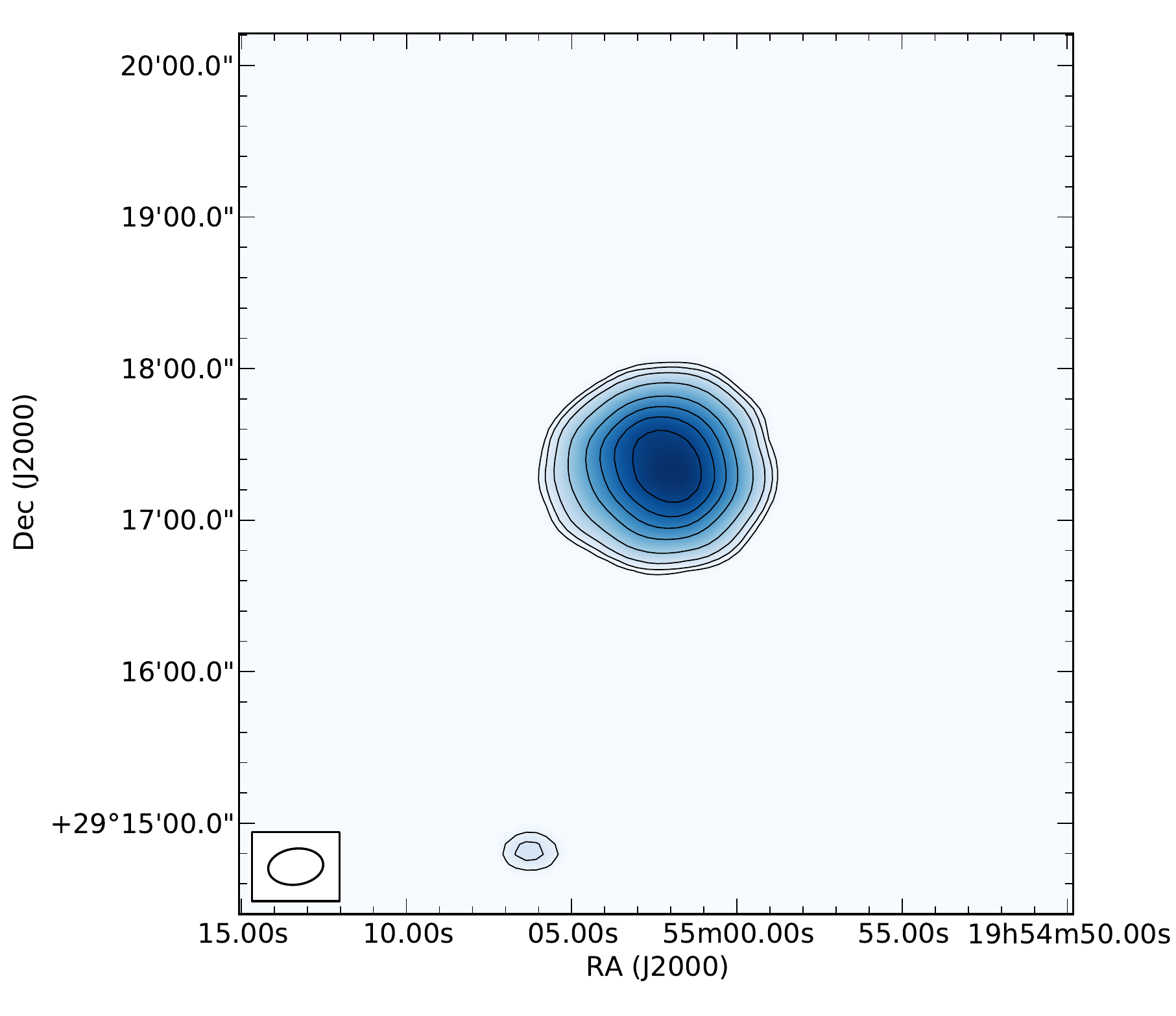}
\caption{VLA image of NGC 6842 at $5\um{GHz}$.}
\label{fig:3558_highres}
\end{center}
\end{figure}

In Figure \ref{fig:3558_highres} we present a 5-GHz map obtained with the VLA in configuration D. So far, it is the second interferometric map at this frequency after the one published by \citet{Zijlstra1989}. The high sensitivity of our map allowed us to determine an accurate flux density of $37.9\pm1.1\um{mJy}$, in good agreement with the single dish estimate of \citet{Gregory1996} of $39\pm5\um{mJy}$. The statistical distance determined from the 5-GHz flux density is $2.2\pm0.3\um{kpc}$.

We used the 5-GHz map to calculate the ionized mass of the nebula. Considering an electron temperature $T_e=13\ 200\um{K}$ after \citet{Kaler1983} the ionized mass of NGC 6842 results of order of $0.1M_\odot$ (with the greatest uncertainties deriving from the distance assumption). Note that \citet{Lenzuni1989} calculated an ionized mass $M_\mathrm{ion}=0.17M_\odot$ and \citet{Gorny1997} for the central star $M_\star\simeq0.6M_\odot$.

\subsubsection{PN G040.3-00.4} 
PN G040.3-00.4 was first detected and recognised as PN by Abell (1955; 1966). In the optical it appears as a ring nebula with a diameter of about $50\um{arcsec}$. Several studies indicates that this PN is located at a distance $d\!\sim\!2\um{kpc}$, though the various estimates range from $1.6$ to $3\um{kpc}$ (\citealt{Cahn1971}; \citealt{Kaler1983}; \citealt{Maciel1984}; \citealt{Sabbadin1986}; \citealt{Cahn1992}; \citealt{Stanghellini1993}; \citealt{Zhang1995}; \citealt{Phillips2004}; \citealt{Ortiz2013}).

The source has been detected at IR wavelengths with \textit{Spitzer} (\citealt{Phillips2008}; \citealt{Kwok2008}) and \textit{AKARI} \citep{Phillips2011}. It was first observed at radio wavelength by \citet{Milne1982} at $14.7\um{GHz}$, then by \citet{Condon1998b} at $1.4\um{GHz}$ (NVSS) and by \citet{Pazderska2009} at $30\um{GHz}$.

\begin{figure*}
\begin{center}
\includegraphics[height=6.7cm]{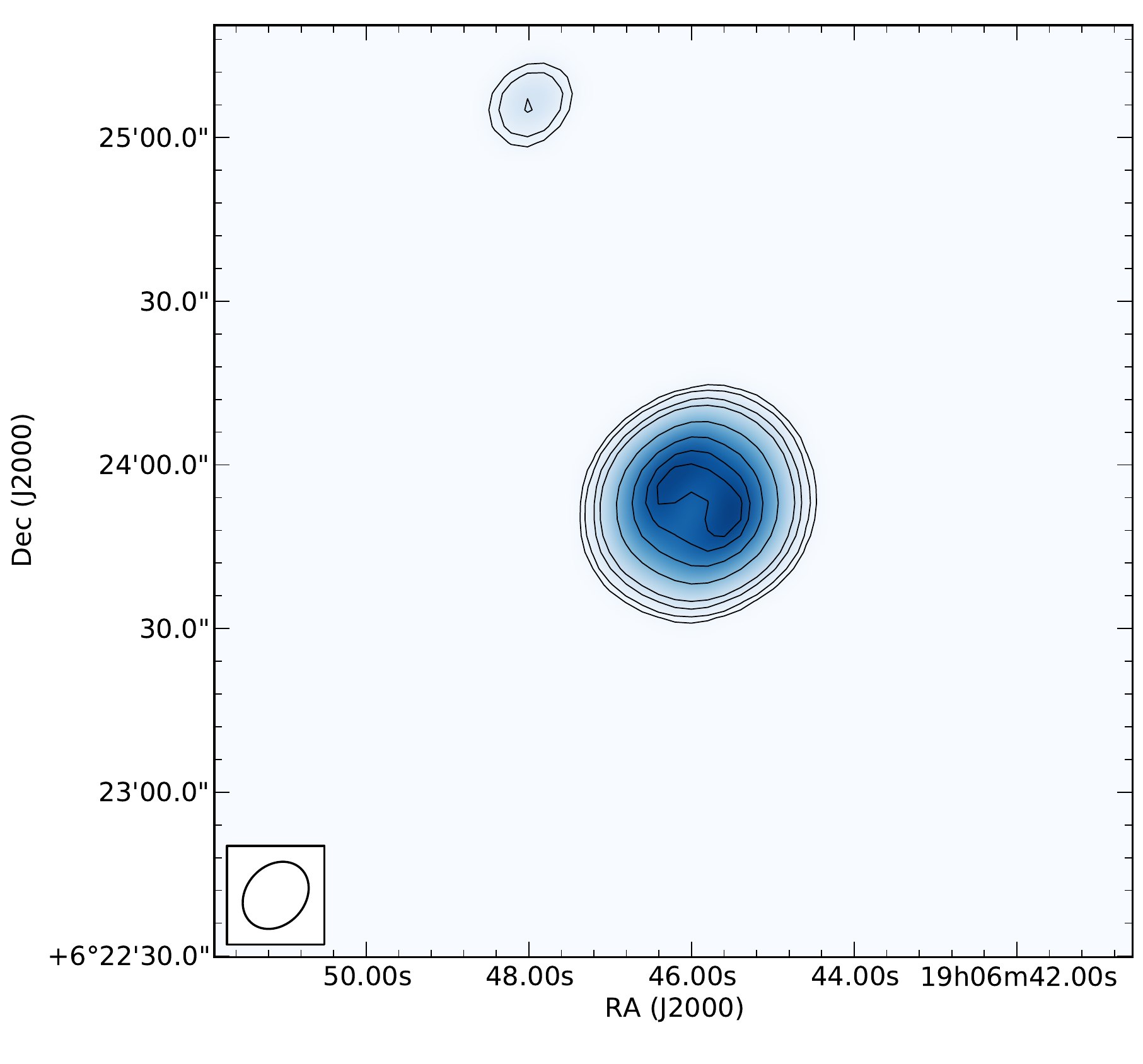}\hspace{1cm}
\includegraphics[height=6.7cm]{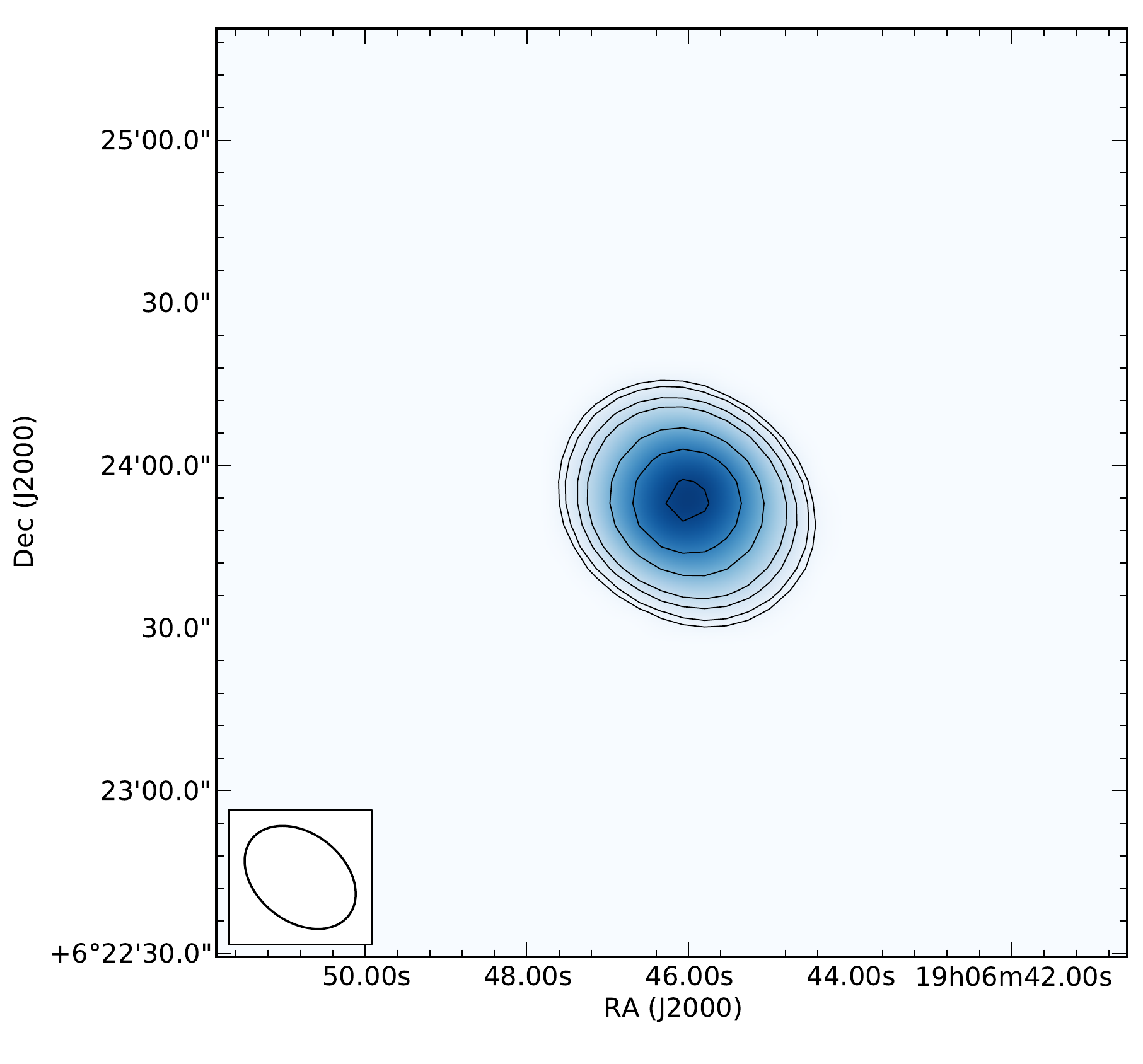}
\caption{VLA image of PN G040-00.4 at $1.4\um{GHz}$ (left) and $5\um{GHz}$ (right).}
\label{fig:3654_highres}
\end{center}
\end{figure*}

We observed the source with the VLA both at $1.4\um{GHz}$ (configuration CnB) and at $5\um{GHz}$ (configuration D). In Figure \ref{fig:3654_highres} we present the two maps. The flux density we computed at $5\um{GHz}$ is $S_C=59.7\pm0.5\um{mJy}$ while at $1.4\um{GHz}$ we find $S_L=64.1\pm0.8\um{mJy}$ and a resulting spectral index $\alpha=-0.06\pm0.01$ (see Paper I). It is important to notice that the NVSS catalogue lists a flux density of $33.6\pm1.1\um{mJy}$ at $1.4\um{GHz}$, significantly lower than our value. A cross check with the VLA Galactic Plane Survey maps \citep{Stil2006}, observed in 2000, revealed a flux density in perfect agreement with our estimate rather than the NVSS one. We conclude that the reported NVSS flux density is very likely incorrect.

From the 5-GHz map we computed a statistical distance of $2.1\pm0.3\um{kpc}$. Assuming an electron temperature $T_e=10^4\um{K}$ (to our knowledge, there is no real estimates of $T_e$ in literature), we can calculate an ionized mass $M_\mathrm{ion}\!\sim\!0.2M_\odot$.

\subsubsection{PN G031.9-00.3}  
\label{sec:3706}
Very little is known about PN G031.9-00.3. Discovered and classified by \citet{Weinberger1981}, a tentative distance of $4.7\um{kpc}$ has been given by \citet{Dobrincic2008} from Galactic rotation curves. It was detected as radio source at $1.4\um{GHz}$ in NVSS and by \citet{White2005} at $5\um{GHz}$.

\begin{figure*}
\begin{center}
\includegraphics[height=6.7cm]{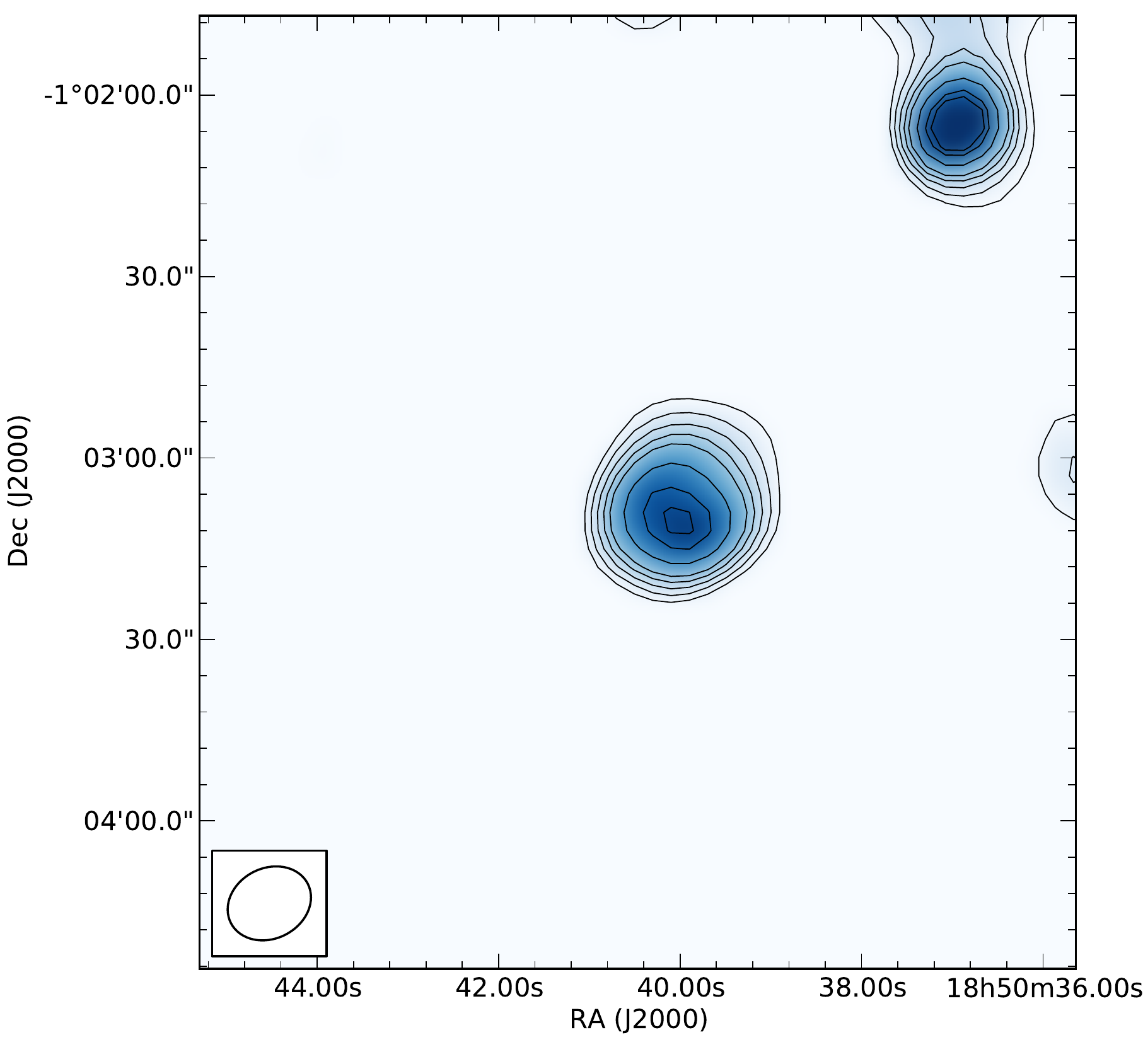}\hspace{1cm}
\includegraphics[height=6.7cm]{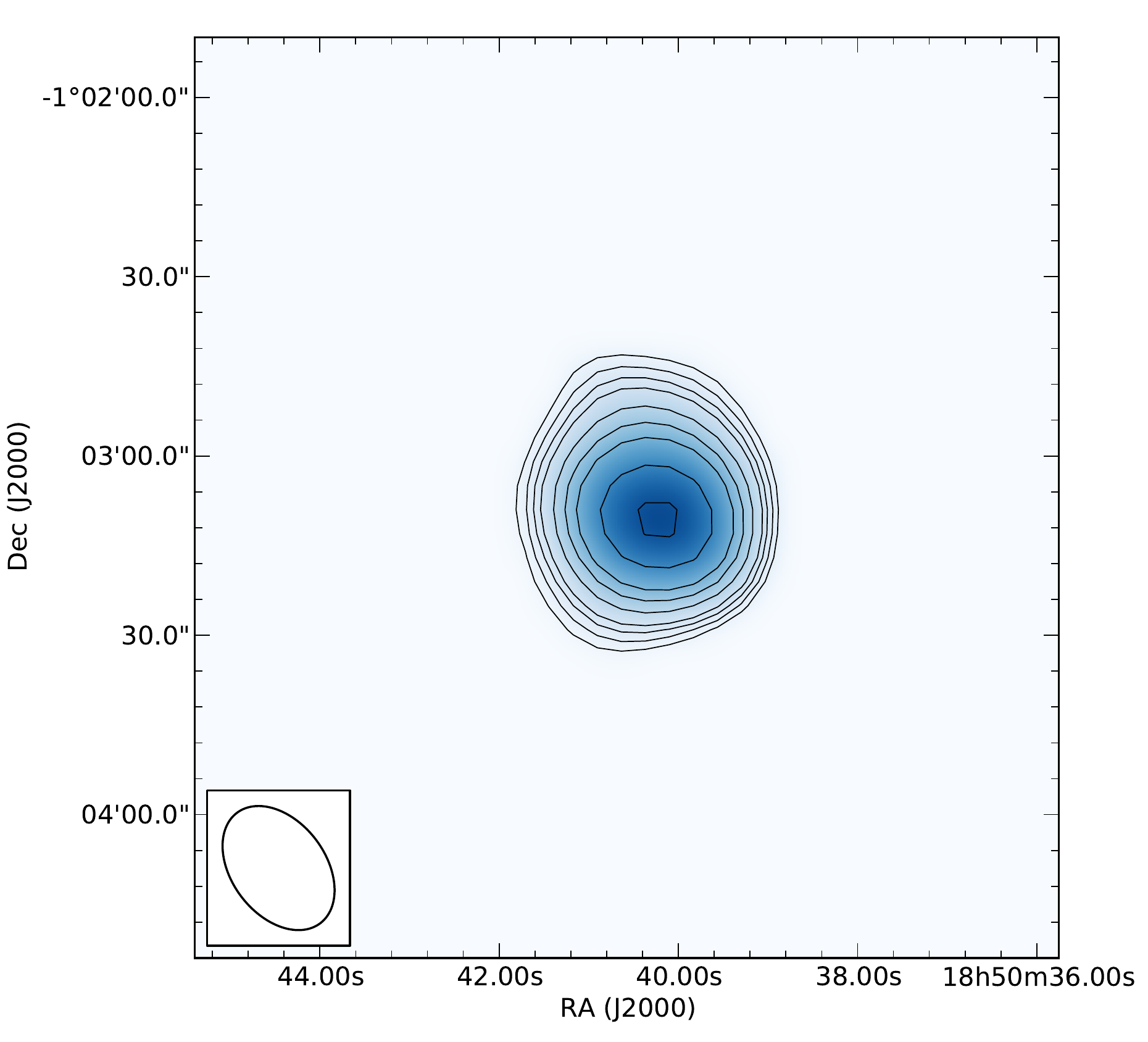}
\caption{VLA image of PN G031.9-00.3 at $1.4\um{GHz}$ (left) and $5\um{GHz}$ (right).}
\label{fig:3706_highres}
\end{center}
\end{figure*}

In Figure \ref{fig:3706_highres} we present the two maps derived from our data. For this PN we found at $5\um{GHz}$ a flux density of $S_C=19.6\pm0.8\um{mJy}$ while at $1.4\um{GHz}$ $S_L=10.8\pm3.7\um{mJy}$ (see Paper I). The resulting spectral index is $\alpha=0.49\pm0.28$, showing that the source is still optically thick at $1.4\um{GHz}$. The statistical distance computed for this PN is $3.2\pm0.5\um{kpc}$, significantly different from the literature value. It is possible that the distance calculation is biased by the fact that the nebula could be not completely optically thin at $5\um{GHz}$ as suggested by the spectral index. For this reason we safely report the ionized mass derived from the literature distance and from our estimate. Assuming an electron temperature $T_e=10^4\um{K}$, we find $M_\mathrm{ion}\!\sim\!0.3M_\odot$ if $D=4.7\um{kpc}$ and $M_\mathrm{ion}\!\sim\!0.1M_\odot$ if $D=3.2\um{kpc}$.

\section{Summary and conclusions}
Radio interferometer observations are an important instrument in the characterisation of Galactic circumstellar envelopes. Not only they permit us to reliably trace the ionized medium surrounding an evolved star, but also to derive different physical parameters of these objects.

However, the synthesis imaging process from interferometer data is optimised for point source observations. When extended sources are to be imaged we found that different algorithms, like the multi-scale CLEAN, give more satisfactory results in terms of image quality. For sources extended more than the interferometer LAS we also complemented the interferometer data with single-dish maps in order to get to a qualitatively and quantitatively reliable representation of the sky.

The combination process, via feathering, resulted fundamental for the two SNRs, for which the mere VLA maps were extremely poor and the recovered flux density was a factor 3 or 4 less than the real one. The information availability in a very wide range of spatial frequencies allowed us to create accurate and detailed images of these objects. Furthermore, the two SNRs served as a test to verify that this simple procedure can be used to perform also quantitative analyses on very extended sources.

For this reason we used SNR G011.2-00.3 mainly to test the method, and in particular the total flux density recovery. This source, in fact, has been deeply studied so far and a good base knowledge is available in literature. What we found is a perfect agreement between our flux density determination and the literature estimates from single-dish data. The resulting map at $5\um{GHz}$ is currently the best representation of the source at this frequency.

For SNR G027.4+00.0 we were able to obtain important results. Using the 1.4-GHz map we posed a more stringent upper limit for the flux density of its yet non-detected pulsar. We did also show that the nebula presents a spatial variation of its radio spectral index. One hypothesis is that the flatter spectral index characterising the western region may be due to inhomogeneities of the strong magnetic field from the central object, causing a superposition of different `regular' synchrotron contributions with different turnover frequencies. This, in turn, results in a global flatter spectrum. Alternatively it is also possible that the spectrum flattening could be ascribed to the impact of the SN shock on dense molecular clouds.

The possibility to accurately measure the flux densities also allowed us to constraint some physical parameters for the four PNe. In particular we were able to determine their ionized masses and their statistical distance. We found that all the four PNe have an ionized mass of order of $0.1M_\odot$. For two of them it is possible to find in literature ionized mass estimates, via $\mathrm{H}\alpha$ or $\mathrm{H}\beta$ fluxes, that are in agreement with our results from radio. For the other two, as far as we know, our values are the first estimates.

We remark that this work has shown the importance of reliable radio data in multiwavelength studies. In fact radio maps can be successfully used to derive important physical parameters for different types of Galactic sources, regardless of their angular extensions. Eventually we want to stress that the procedure adopted for interferometer and single-dish data combination does not depend on the particular instruments used (VLA and GBT), since it works directly on FITS images. Therefore, given the positive results showed in this work, it can be applied as it is to different other interferometers such as ALMA, where flux-loss issues are even more important, and, in future, SKA.

\section*{Acknowledgements}
This work is based on observations made with the Very Large Array and the Green Bank Telescope of the National Radio Astronomy Observatory, a facility of the National Science Foundation operated under cooperative agreement by Associated Universities Inc.. Archive search made use of the SIMBAD database and the VizieR catalogue access tool, operated by the Centre de Donn\'ees astronomique de Strasbourg.


\begin{thebibliography}{99}
\bibitem[\protect\citeauthoryear{Abell}{1955}]{Abell1955} Abell G.~O., 1955, PASP, 67, 25
\bibitem[\protect\citeauthoryear{Abell}{1966}]{Abell1966} Abell G.~O., 1966, ApJ, 144, 259 
\bibitem[\protect\citeauthoryear{Altenhoff et al.}{1970}]{Altenhoff1970} Altenhoff W.~J., Downes D., Goad L., Maxwell A., Rinehart R., 1970, A\&AS, 1, 319
\bibitem[\protect\citeauthoryear{Angerhofer, Becker \& Kundu}{1977}]{Angerhofer1977} Angerhofer P.~E., Becker R.~H., Kundu M.~R., 1977, A\&A, 55, 11
\bibitem[\protect\citeauthoryear{Bhatnagar et al.}{2011}]{Bhatnagar2011} Bhatnagar S., Rau U., Green D.~A., Rupen M.~P., 2011, ApJ, 739, L20
\bibitem[\protect\citeauthoryear{Becker \& Kundu}{1975}]{Becker1975} Becker R.~H., Kundu M.~R., 1975, AJ, 80, 679 
\bibitem[\protect\citeauthoryear{Cahn \& Kaler}{1971}]{Cahn1971} Cahn J.~H., Kaler J.~B., 1971, ApJS, 22, 319
\bibitem[\protect\citeauthoryear{Cahn \& Rubin}{1974}]{Cahn1974} Cahn J.~H., Rubin R.~H., 1974, AJ, 79, 128
\bibitem[\protect\citeauthoryear{Cahn, Kaler \& Stanghellini}{1992}]{Cahn1992} Cahn J.~H., Kaler J.~B., Stanghellini L., 1992, A\&AS, 94, 399 
\bibitem[\protect\citeauthoryear{Carey et al.}{2009}]{Carey2009} Carey S.~J., et al., 2009, PASP, 121, 76
\bibitem[\protect\citeauthoryear{Caswell et al.}{1982}]{Caswell1982} Caswell J.~L., Haynes R.~F., Milne D.~K., Wellington K.~J., 1982, MNRAS, 200, 1143 
\bibitem[\protect\citeauthoryear{Clark}{1980}]{Clark1980} Clark B.~G., 1980, A\&A, 89, 377
\bibitem[\protect\citeauthoryear{Clark}{1999}]{Clark1999} Clark B.~G., 1999, in Taylor G.~B., Carilli C.~L., Perley R.~A., eds, ASP Conf. Ser. Vol. 180, Synthesis imaging in radio astronomy II. Astron. Soc. Pac., San Francisco, p. 1
Scappini F., Caselli P., Attolini M.~R., 1995, MNRAS, 276, 57 
\bibitem[\protect\citeauthoryear{Condon \& Kaplan}{1998}]{Condon1998b} Condon J.~J., Kaplan D.~L., 1998, ApJS, 117, 361 
\bibitem[\protect\citeauthoryear{Condon et al.}{1998}]{Condon1998} Condon J.~J., Cotton W.~D., Greisen E.~W., Yin Q.~F., Perley R.~A., Taylor G.~B., Broderick J.~J., 1998, AJ, 115, 1693 
\bibitem[\protect\citeauthoryear{Cornwell}{2008}]{Cornwell2008} Cornwell T.~J., 2008, IEEE Journal of Selected Topics in Signal Processing, Vol. 2, Issue 5, p.793-801
\bibitem[\protect\citeauthoryear{Curtis}{1919}]{Curtis1919} Curtis H.~D., 1919, PASP, 31, 285 
\bibitem[\protect\citeauthoryear{Dickel \& Milne}{1972}]{Dickel1972} Dickel J.~R., Milne D.~K., 1972, AuJPh, 25, 539
\bibitem[\protect\citeauthoryear{Dobrin{\v c}i{\'c} et al.}{2008}]{Dobrincic2008} Dobrin{\v c}i{\'c} M., Villaver E., Guerrero M.~A., Manchado A., 2008, AJ, 135, 2199 
\bibitem[\protect\citeauthoryear{Downes}{1984}]{Downes1984} Downes A., 1984, MNRAS, 210, 845
\bibitem[\protect\citeauthoryear{Dreyer}{1888}]{Dreyer1888} Dreyer J.~L.~E., 1888, MmRAS, 49, 1
\bibitem[\protect\citeauthoryear{Emerson \& Gr\"ave}{1988}]{Emerson1988} Emerson D.~T., Gr\"ave R., 1988, A\&A, 190, 353
\bibitem[\protect\citeauthoryear{Frew \& Parker}{2010}]{Frew2010} Frew D.~J., Parker Q.~A., 2010, Publ. Astron. Soc. Aust., 27, 129 
\bibitem[\protect\citeauthoryear{Frew et al.}{2014}]{Frew2014} Frew D.~J., et al., 2014, MNRAS, 440, 1345 
\bibitem[\protect\citeauthoryear{Gaensler \& Slane}{2006}]{Gaensler2006} Gaensler B.~M., Slane P.~O., 2006, ARA\&A, 44, 17 \bibitem[\protect\citeauthoryear{Giacani et al.}{2011}]{Giacani2011} Giacani E., Smith M.~J.~S., Dubner G., Loiseau N., 2011, A\&A, 531, A138
\bibitem[\protect\citeauthoryear{Giammanco et al.}{2011}]{Giammanco2011} Giammanco C., et al., 2011, A\&A, 525, A58 
\bibitem[\protect\citeauthoryear{Glushak}{2014}]{Glushak2014} Glushak A.~P., 2014, ARep, 58, 6 
\bibitem[\protect\citeauthoryear{Gorny, Stasi{\'n}ska \& Tylenda}{1997}]{Gorny1997} Gorny S.~K., Stasi{\'n}ska G., Tylenda R., 1997, A\&A, 318, 256
\bibitem[\protect\citeauthoryear{Gotthelf \& Vasisht}{1997}]{Gotthelf1997} Gotthelf E.~V., Vasisht G., 1997, ApJ, 486, L133
\bibitem[\protect\citeauthoryear{Gotthelf, Vasisht \& Dotani}{1999}]{Gotthelf1999} Gotthelf E.~V., Vasisht G., Dotani T., 1999, ApJ, 522, L49
\bibitem[\protect\citeauthoryear{Gregory \& Taylor}{1986}]{Gregory1986} Gregory P.~C., Taylor A.~R., 1986, AJ, 92, 371
\bibitem[\protect\citeauthoryear{Gregory et al.}{1996}]{Gregory1996} Gregory P.~C., Scott W.~K., Douglas K., Condon J.~J., 1996, ApJS, 103, 427
\bibitem[\protect\citeauthoryear{Helfand et al.}{2006}]{Helfand2006} Helfand D.~J., Becker R.~H., White R.~L., Fallon A., Tuttle S., 2006, AJ, 131, 252
\bibitem[\protect\citeauthoryear{H\"ogbom}{1974}]{Hogbom1974} H\"ogbom J.A., 1974, A\&AS, 15, 417
\bibitem[\protect\citeauthoryear{Hromov \& Kohomek}{1968}]{Hromov1968} Hromov G.~S., Kohomek L., 1968, BAICz, 19, 1
\bibitem[\protect\citeauthoryear{Ingallinera et al.}{2014}]{Ingallinera2014} Ingallinera A., et al., 2014, MNRAS, 437, 3626
\bibitem[\protect\citeauthoryear{Kaler}{1983}]{Kaler1983} Kaler J.~B., 1983, ApJ, 271, 188 
\bibitem[\protect\citeauthoryear{Kassim}{1988}]{Kassim1988} Kassim N.~E., 1988, ApJS, 68, 715 
\bibitem[\protect\citeauthoryear{Kassim}{1989}]{Kassim1989} Kassim N.~E., 1989, ApJ, 347, 915 
\bibitem[\protect\citeauthoryear{Kriss et al.}{1985}]{Kriss1985} Kriss G.~A., Becker R.~H., Helfand D.~J., Canizares C.~R., 1985, ApJ, 288, 703 
\bibitem[\protect\citeauthoryear{Koo \& Heiles}{1991}]{Koo1991} Koo B.-C., Heiles C., 1991, ApJ, 382, 204
\bibitem[\protect\citeauthoryear{Koo et al.}{2007}]{Koo2007} Koo B.-C., Moon D.-S., Lee H.-G., Lee J.-J., Matthews K., 2007, ApJ, 657, 308 
\bibitem[\protect\citeauthoryear{Kothes \& Reich}{2001}]{Kothes2001} Kothes R., Reich W., 2001, A\&A, 372, 627 
\bibitem[\protect\citeauthoryear{Kumar \& Safi-Harb}{2010}]{Kumar2010} Kumar H.~S., Safi-Harb S., 2010, ApJ, 725, L191
\bibitem[\protect\citeauthoryear{Kumar et al.}{2014}]{Kumar2014} Kumar H.~S., Safi-Harb S., Slane P.~O., Gotthelf E.~V., 2014, ApJ, 781, 41 
\bibitem[\protect\citeauthoryear{Kwok et al.}{2008}]{Kwok2008} Kwok S., Zhang Y., Koning N., Huang H.-H., Churchwell E., 2008, ApJS, 174, 426 
\bibitem[\protect\citeauthoryear{Large, Mathewson \& Haslam}{1961}]{Large1961} Large M.~I., Mathewson D.~S., Haslam C.~G.~T., 1961, MNRAS, 123, 123
\bibitem[\protect\citeauthoryear{Lenzuni, Natta \& Panagia}{1989}]{Lenzuni1989} Lenzuni P., Natta A., Panagia N., 1989, ApJ, 345, 306 
\bibitem[\protect\citeauthoryear{Lin et al.}{2011}]{Lin2011} Lin L., et al., 2011, ApJ, 740, L16
\bibitem[\protect\citeauthoryear{Maciel}{1984}]{Maciel1984} Maciel W.~J., 1984, A\&AS, 55, 253
\bibitem[\protect\citeauthoryear{McMullin et al.}{2007}]{McMullin2007} McMullin J.~P., Waters B., Schiebel D., Young W., Golap K., 2007, ASPC, 376, 127
\bibitem[\protect\citeauthoryear{Milne}{1969}]{Milne1969} Milne D.~K., 1969, AuJPh, 22, 613 
\bibitem[\protect\citeauthoryear{Milne}{1979}]{Milne1979} Milne D.~K., 1979, AuJPh, 32, 83
\bibitem[\protect\citeauthoryear{Milne \& Aller}{1982}]{Milne1982} Milne D.~K., Aller L.~H., 1982, A\&AS, 50, 209
\bibitem[\protect\citeauthoryear{Mizuno et al.}{2010}]{Mizuno2010} Mizuno D.~R., et al., 2010, AJ, 139, 1542
\bibitem[\protect\citeauthoryear{Morsi \& Reich}{1987}]{Morsi1987} Morsi H.~W., Reich W., 1987, A\&AS, 71, 189
\bibitem[\protect\citeauthoryear{O'dell}{1962}]{Odell1962} O'dell C.~R., 1962, ApJ, 135, 371 
\bibitem[\protect\citeauthoryear{Ortiz}{2013}]{Ortiz2013} Ortiz R., 2013, A\&A, 560, A85 
\bibitem[\protect\citeauthoryear{Pazderska et al.}{2009}]{Pazderska2009} Pazderska B.~M., et al., 2009, A\&A, 498, 463
\bibitem[\protect\citeauthoryear{Phillips}{2004}]{Phillips2004} Phillips J.~P., 2004, MNRAS, 353, 589
\bibitem[\protect\citeauthoryear{Phillips \& Ramos-Larios}{2008}]{Phillips2008} Phillips J.~P., Ramos-Larios G., 2008, MNRAS, 383, 1029 
\bibitem[\protect\citeauthoryear{Phillips \& M{\'a}rquez-Lugo}{2011}]{Phillips2011} Phillips J.~P., M{\'a}rquez-Lugo R.~A., 2011, RMxAA, 47, 83 
\bibitem[\protect\citeauthoryear{Reynolds et al.}{1994}]{Reynolds1994} Reynolds S.~P., Lyutikov M., Blandford R.~D., Seward F.~D., 1994, MNRAS, 271, L1 
\bibitem[\protect\citeauthoryear{Rieke et al.}{2004}]{Rieke2004} Rieke G.~H., et al., 2004, ApJS, 154, 25
\bibitem[\protect\citeauthoryear{Sabbadin}{1986}]{Sabbadin1986} Sabbadin F., 1986, A\&A, 160, 31
\bibitem[\protect\citeauthoryear{Sanbonmatsu \& Helfand}{1992}]{Sanbonmatsu1992} Sanbonmatsu K.~Y., Helfand D.~J., 1992, AJ, 104, 2189 
\bibitem[\protect\citeauthoryear{Shaver \& Goss}{1970}]{Shaver1970} Shaver P.~A., Goss W.~M., 1970, AuJPA, 14, 77
\bibitem[\protect\citeauthoryear{Stanghellini, Corradi, \& Schwarz}{1993}]{Stanghellini1993} Stanghellini L., Corradi R.~L.~M., Schwarz H.~E., 1993, A\&A, 279, 521
\bibitem[\protect\citeauthoryear{Stil et al.}{2006}]{Stil2006} Stil J.~M., et al., 2006, AJ, 132, 1158
\bibitem[\protect\citeauthoryear{Tajitsu \& Tamura}{1998}]{Tajitsu1998} Tajitsu A., Tamura S., 1998, AJ, 115, 1989
\bibitem[\protect\citeauthoryear{Taylor \& Gregory}{1983}]{Taylor1983} Taylor A.~R., Gregory P.~C., 1983, AJ, 88, 1784
\bibitem[\protect\citeauthoryear{Taylor et al.}{1996}]{Taylor1996} Taylor A.~R., Goss W.~M., Coleman P.~H., van Leeuwen J., Wallace B.~J., 1996, ApJS, 107, 239
\bibitem[\protect\citeauthoryear{Tian \& Leahy}{2008}]{Tian2008} Tian W.~W., Leahy D.~A., 2008, ApJ, 677, 292
\bibitem[\protect\citeauthoryear{Todt et al.}{2013}]{Todt2013} Todt H., et al., 2013, MNRAS, 430, 2302 
\bibitem[\protect\citeauthoryear{Torii et al.}{1997}]{Torii1997} Torii K., Tsunemi H., Dotani T., Mitsuda K., 1997, ApJ, 489, L145 
\bibitem[\protect\citeauthoryear{van de Steene \& Zijlstra}{1995}]{VandeSteene1995} van de Steene G.~C., Zijlstra A.~A., 1995, A\&A, 293, 541
\bibitem[\protect\citeauthoryear{Vasisht et al.}{1996}]{Vasisht1996} Vasisht G., Aoki T., Dotani T., Kulkarni S.~R., Nagase F., 1996, ApJ, 456, L59
\bibitem[\protect\citeauthoryear{Vasisht \& Gotthelf}{1997}]{Vasisht1997} Vasisht G., Gotthelf E.~V., 1997, ApJ, 486, L129 
\bibitem[\protect\citeauthoryear{Wachter et al.}{2010}]{Wachter2010} Wachter S., Mauerhan J.~C., Van Dyk S.~D., Hoard D.~W., Kafka S., Morris P.~W., 2010, AJ, 139, 2330
\bibitem[\protect\citeauthoryear{Weinberger \& Sabbadin}{1981}]{Weinberger1981} Weinberger R., Sabbadin F., 1981, A\&A, 100, 66
\bibitem[\protect\citeauthoryear{White, Becker \& Helfand}{2005}]{White2005} White R.~L., Becker R.~H., Helfand D.~J., 2005, AJ, 130, 586
\bibitem[\protect\citeauthoryear{Wilson, Rohlfs \& Huttemeister}{2012}]{Wilson2012} Wilson T.~L., Rohlfs K., Huttemeister S., 2012, Tools of Radio Astronomy, fifth edition, Springer
\bibitem[\protect\citeauthoryear{Zhang}{1995}]{Zhang1995} Zhang C.~Y., 1995, ApJS, 98, 659
\bibitem[\protect\citeauthoryear{Zijlstra, Pottasch \& Bignell}{1989}]{Zijlstra1989} Zijlstra A.~A., Pottasch S.~R., Bignell C., 1989, A\&AS, 79, 329

\end{thebibliography}
\end{document}